\documentclass{article}
\usepackage{amsfonts}
\usepackage{amsmath, amssymb}
\usepackage{graphicx}
\usepackage{color}
\usepackage{ulem}
\usepackage{epsfig}

\newtheorem{theorem}{Theorem}

\newtheorem{definition}[theorem]{Definition}

\newtheorem{remark}[theorem]{Remark}

\begin{document}

\title{Distinguished trajectories in time dependent vector fields}

\author{J. A. Jim\'enez Madrid, Ana M. Mancho\\
Instituto de Ciencias Matem\'aticas, CSIC-UAM-UC3M-UCM, \\
Serrano 121, 28006 Madrid (Spain)}


\maketitle

\begin{abstract}

We introduce a new definition of distinguished trajectory that
generalises the concepts of fixed point and periodic orbit 
to aperiodic dynamical
systems. This new definition is valid for identifying 
distinguished trajectories with hyperbolic and non-hyperbolic types 
of stability. 
The definition is implemented numerically  and the procedure 
consist in determining a path of {\it limit coordinates}.
It has been successfully applied to known examples of distinguished trajectories.
In the context of highly aperiodic realistic flows our definition
characterises {\it distinguished trajectories} in finite  time intervals,
and states that outside these intervals trajectories are no longer {\it distinguished}.

\end{abstract}

\baselineskip 20pt

\noindent Corresponding author: Ana M. Mancho,  Instituto de Ciencias Matem\'aticas, 
Consejo Superior de Investigaciones Cient\'{\i}ficas,
Serrano 121, 28006, Madrid. e-mail: A.M.Mancho@imaff.cfmac.csic.es.
Phone: +34 91 5616800 ext 2408. Fax: +34 91
5854894. 

\baselineskip 20pt \newpage

{\bf This paper  attempts to generalise the concepts of fixed point and periodic orbit
to time dependent aperiodic dynamical systems. Fixed points and periodic orbits are  keystones
for describing  solutions of autonomous 
and time periodic dynamical systems, as the stable and unstable manifolds of these hyperbolic objects
 form  the basis of the 
geometrical template organising the description of  the dynamical system.
The mathematical theory  of aperiodic dynamical systems is far from  complete. In this context, this work
deals with a general definition that encompasses the concepts of fixed point
and periodic orbit and which when  applied  to finite time and 
aperiodic dynamical systems identifies special trajectories that play an 
organising role in  the geometry of the flow. }


\section{Introduction}
 In recent years the theory of dynamical systems 
has provided a useful framework for describing transport
in fluid  flows. Since the seminal work 
by Aref \cite{Aref} on chaotic advection
much progress has been made both in theory and applications.
Dynamical systems techniques were first applied to Lagrangian transport in the context of two-dimensional, time-periodic flows
\cite{tpfl} and stationary 3D flows such
as the ABC flow \cite{3dsfl}. 
More recently these techniques have been extended to describe aperiodic flows \cite{wiggins,malhotra,haller1} and finite
time-dependent flows, such 
as those rising in geophysical applications \cite{physrep,nlpg}.
However, the mathematical theory for both aperiodic time-dependent flows and finite time aperiodic flows
 is far from being  completely developed.

For stationary flows the idea of {\it fixed point} is a key for describing
geometrically the solutions. Fixed points may be classified as hyperbolic or non-hyperbolic 
depending on their stability  properties. 
Stable and unstable manifolds 
of hyperbolic fixed points organise the phase portraits of the flow away from the region close to the fixed points \cite{wigg,guck}.
These manifolds comprise respectively  the trajectories that 
approach the fixed points as time tends to plus or minus infinite. As 
they are formed of trajectories they act as barriers to transport 
as particles cannot cross them without violating the
uniqueness of the solution. They are useful because 
they allow qualitative predictions for the evolution of sets of initial conditions avoiding  
explicit integration of 
initial conditions on the whole domain. 
Hyperbolic fixed points and their stable and unstable manifolds are the basic notions used for 
the geometrical description of flows in autonomous dynamical system.

The concept of fixed point is extended to time periodic flows by means of the Poincar\'e map, as periodic orbits
with  period $T$ become fixed points of 
the Poincar\'e map. For hyperbolic periodic orbits there exist also stable and unstable 
manifolds that are geometric objects that organise the global dynamics. 
Again they are respectively the sets of orbits asymptotically 
approaching the periodic orbit as time tends to plus or minus infinity.

Aperiodic flows are still poorly understood, as theory that
is well established for autonomous or periodic flows does not apply  to them directly. 
For instance  there exists  efforts in the mathematical community to extend the well known concept 
of bifurcation for stationary flows  to non-autonomous systems \cite{langa1,langa2}.
To gain insight on the geometrical structure of  aperiodic flows, concepts such as Lyapunov exponents  
are used, however these are  defined strictly on infinite time systems. 
Realistic flows, like those arising  in geophysics or oceanography, 
are not infinite time systems and for their description,  finite time versions of 
the definition of Lyapunov exponents such as 
 Finite Size Lyapunov Exponents (FSLE) \cite{aurell} and 
Finite Time Lyapunov Exponents (FTLE)  \cite{haller,nese} are used.
Special trajectories, such as
detachment and reattachment points   \cite{haller2},
are observed  in  highly aperiodic  or turbulent flows.
In particular these separation trajectories occur on the boundaries 
in simplified ocean models \cite{coulliete} and
also in technological applications in air foil design \cite{eis}.
Recent articles by Ide {\it et al} and Ju {\it et al} \cite{kayo,yu} referring 
to these special trajectories
introduce the concept of {\it Distinguished Hyperbolic Trajectory } (DHT) which encompases 
not only trajectories on the boundaries but also
special trajectories in the interior of the flow. DHT are
hyperbolic trajectories that, like hyperbolic  fixed points and periodic orbits,
have  stable and unstable manifolds that 
are key for describing geometrically the solutions on the phase space.
This generalisation is an important step-forwards in the study of aperiodic flows, as it is a powerful tool
for describing transport in realistic oceanographic  flows \cite{physrep, nlpg, jpo,br}.
Distinguished hyperbolic trajectories as defined in \cite{kayo,yu}
are computed from hyperbolic instantaneous stagnation points (ISPs)
by means of an iterative procedure. If 
instantaneous stagnation points bifurcate and do not persist for all times
the technique  developed in \cite{kayo,yu} cannot be applied in those time intervals, leaving many questions  unanswered,
such as what happens to the distinguished trajectories at those times, 
for 
distinguished hyperbolic trajectories
are trajectories, and as trajectories  exist at all times. 
In fact  Ref. \cite{physrep} provides examples of  vector fields
with exact distinguished hyperbolic trajectories that exist on  time intervals without hyperbolic ISP. 
Refs. \cite{physrep,nlpg}  discuss
 the impossibility of this technique for  tracking DHTs after  ISP
bifurcations and as a consequence the difficulties in establishing whether DHTs 
obtained at different times are part of the same trajectory or not.

In this paper, following  ideas discussed in  \cite{physrep,kayo,yu}, 
we propose a new definition of  {\it Distinguished Trajectory } (DT)  
which generalises the concepts of fixed point and periodic orbit to aperiodic flows.
We have taken the liberty of calling them {\it Distinguished} as in \cite{physrep,kayo,yu}, 
since although the definitions are not strictly 
equivalent, it is found that  the studied hyperbolic  trajectories 
are encompassed by both definitions.
We remark that our notion has the advantage over the method proposed in  \cite{kayo,yu} 
that the DTs may be computed without the presence of hyperbolic instantaneous stagnation points. 
Our definition  does not
depend on the  dimension $n$ of the space on which the vector field is defined and is valid both for hyperbolic and non-hyperbolic types of stabilities.
Non-hyperbolic DTs have not been studied in \cite{kayo,yu}, and in this sense our definition is broader
than that proposed there. In particular, we will show that exact non-hyperbolic periodic orbits  
fall within the category of  distinguished trajectories. 
Trajectories of this type could be of special interest for their applications in oceanography, as they are related to eddies and vortices.
Ocean eddies are well studied \cite{samelson}. Frequently they are
long lived, and water trapped inside can maintain its biogeochemical properties
for long time, being transported with the vortex. In steady horizontal velocity
fields, the presence of closed streamlines is the mathematical reason
for the isolation of the vortex core from the exterior fluid.
In two-dimensional, incompressible, time-periodic velocity fields the KAM tori
enclose {\it the core}, a region of bounded fluid particle motions   that do not mix with the 
surrounding region \cite{wiggins}. But how to define an eddy from the Lagrangian point of
view in aperiodic flows? This is still an open question \cite{br,haller3} for which we  will discuss 
new possibilities suggested by  the definitions given in this article.

The structure of the paper is as follows. Section 2 introduces the definition 
of distinguished trajectory and explains its motivation in the context of 1D examples.
Section 3 explains the algorithm used to verify the applicability of our definition 
of distinguished trajectories to the solutions of the periodically forced Duffing equation.
Details about technical issues arising from implementation of the definition are given.
Section 4 reports the results obtained in several  other 2D and 3D examples, 
both periodic and non-periodic, hyperbolic and non-hyperbolic.
Section 5 discusses  results on realistic flows. Attention is paid to open questions on 
distinguished trajectories  such as those mentioned above and pointed out in Refs. \cite{nlpg,physrep}.
Finally,  section 6 presents the conclusions.

\section{Distinguished trajectories: a definition}

We start by recalling the definition of distinguished hyperbolic trajectory 
 provided  in \cite{kayo}.  Given the system:

\begin{eqnarray}
\frac{d {\bf x}}{dt} &=&{\bf D x} + g^{NL}({\bf x},t) \label{eq1} \, \, {\bf x} \in \, \mathbb{R}^n
\end{eqnarray}

Let ${\bf x}(t)$ be a trajectory of Eq. (\ref{eq1}) that
remains in a bounded region for all time. Then ${\bf x}(t)$ is said to
be a distinguished hyperbolic trajectory if:

  1. it is hyperbolic,

  2. there exists a neighbourhood $\mathcal{B}$ in the flow domain having
     the property that the DHT remains in $\mathcal{B}$ for all time, and
     all other trajectories starting in $\mathcal{B}$ leave $\mathcal{B}$ in finite time,
     as time evolves in either a positive or negative sense,

  3. it is not a hyperbolic trajectory contained in the chaotic
     invariant set created by the intersection of the stable and
     unstable manifolds of another hyperbolic trajectory.

\begin{remark}  If the data spans only a finite time interval, then the DHT
cannot be determined uniquely. Instead, there is a small
region in  ${\mathcal B}$ where the DHT can exist. 
\end{remark}

In \cite{kayo} this setup is extended to general vector fields as follows. Coordinate 
transformations are sought which put the system in the form of Eq. 1 and then 
the previous definition is applied.

We give now our definition of  distinguished trajectory for a general vector field:

\begin{eqnarray}
\frac{d{\bf x}}{dt} &=& {\bf v}({\bf x},t), \label{adv1}  \, \,{\bf x} \in \, \mathbb{R}^n, \, t \in \, \mathbb{R}
\end{eqnarray}
We assume that ${\bf v}({\bf x},t)$ is  $C^r$ ($r \geq 1$) in ${\bf x}$ and continuous in $t$. This will allow for unique solutions
to exist, and also permit linearization, although linearization will not be used in our construction.

Before giving our definition of DT, we first need to introduce some notation and to make some definitions. 
Let ${\bf x}(t)$ denote a trajectory of the system (\ref{adv1}) and 
denote its components in $\mathbb{R}^n$
by $(x_1,x_2,...x_n)$.
For any initial condition ${\bf x^*}$   in an open set ${\mathcal B}\subset\mathbb{R}^n$, 
consider the  function $M: {\mathcal B}\to\mathbb{R}$ 
\begin{equation}
M({\bf x^*})_{t^*,\tau}= \left(  \int^{t^*+\tau}_{t^*-\tau} \! \!\!\sqrt{\sum_{i=1}^n \left(\frac{d x_i(t)}{dt}\right)^2 }dt \right), \label{def:M}
\end{equation}
$M$ is the function that associates to each initial condition ${\bf x^*}$ in ${\mathcal B}$ the arclength of the trajectory that 
passes through ${\bf x^*}$ at time $t^*$. The arclength of the trajectory is considered over its projection in the phase 
space $(x_1,x_2,...x_n)$ and depends on $t^*$ and $\tau$. 
As the function M is defined over an open set it does not necessarily attain a 
minimum, but if it does, the minimum is denoted by  ${\rm min}(M({\bf x^*})_{t^*,\tau})$.

\begin{definition} ($\tau$-Distinguished trajectory).  A trajectory $\gamma(t)$
of Eq. (\ref{adv1}) is $\tau$-distinguished at time $t^*$  if there exists 
an open set ${\mathcal B}$ 
around $\gamma(t^*)$ on which the defined function $M({\bf x^*})_{t^*,\tau}$ has a  minimum 
  and
\begin{eqnarray}
{\rm min}(M({\bf x^*})_{t^*,\tau})=M( \gamma(t^*) )_{t^*,\tau}.
\end{eqnarray}\label{def:dt}
\end{definition}

\subsection{A discussion of the definition}
\begin{figure}[htb]
\begin{center}
\includegraphics[width=1.2\columnwidth]{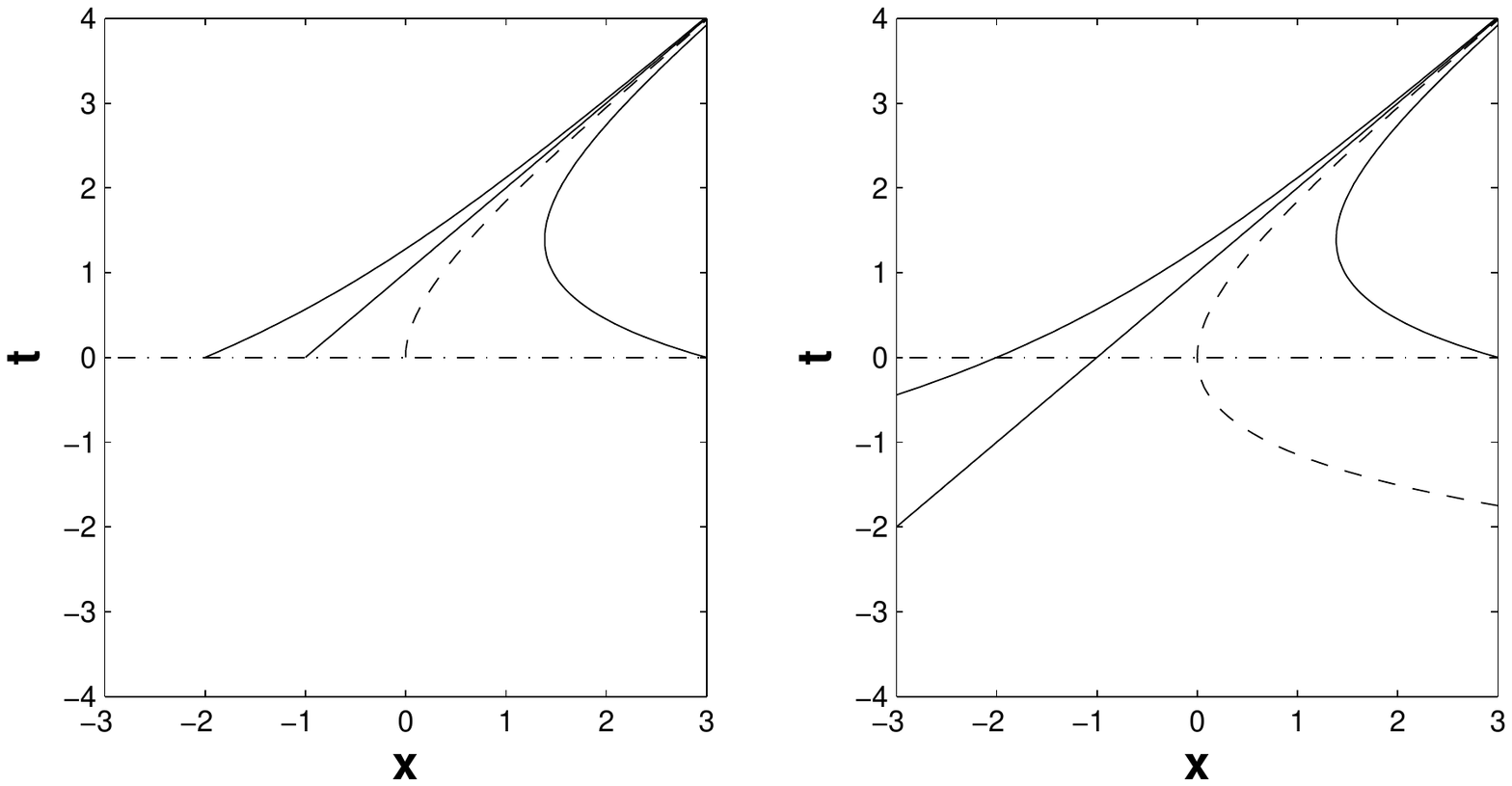}
\end{center}
\vspace{-5cm}
\caption{}\label{fig:1}
\end{figure}

\begin{figure}
\vspace{-5cm}
\begin{center}
\includegraphics[width=1.3\columnwidth]{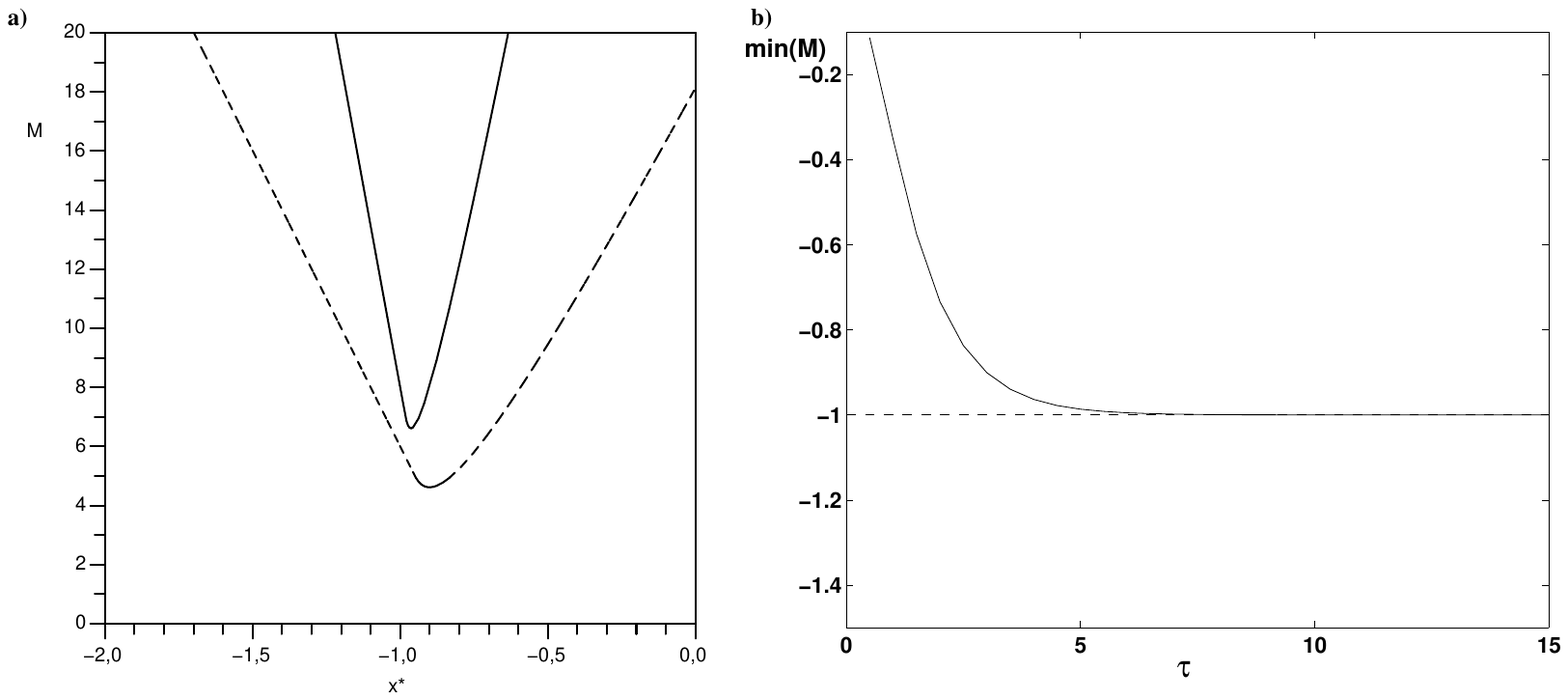}
\end{center}
\caption{}\label{fig:2}
\end{figure}

The
elements of the above definition deserve a detailed justification. 
We illustrate our explanations with examples in 1D. First we consider an example taken from \cite{kayo,szeri}. 
It is the linear one-dimensional non-autonomous dynamical system given by:
\begin{eqnarray}
\frac{dx}{dt}&=&-x+t. \label{eq1b}
\end{eqnarray}
For this example we consider the DHT reported  in \cite{kayo}, which is given by $x=t-1$. This is the particular 
solution of the linear equation (\ref{eq1b}) towards which
all trajectories decay. The solution through the point $x^*$ at $t=0$ is  given by,
\begin{eqnarray}
x(t)=t-1+e^{-t} (x^*+1). \label{sol1}
\end{eqnarray}
Figure \ref{fig:1}a) displays several trajectories starting at 
times ranging from $t = 0$ to $t = 4$ and figure \ref{fig:1}b) displays the same but starting at time 
$t=-4$ (note that in this case part of the trajectories are out of the displayed domain). For each initial 
condition the function $M$ provides the length of the projection of the trajectory  
over the $x$-axis in the range of times $[-\tau, \tau]$.
Geometrically  it is clear that in this example  the function $M$ should have a minimum for a certain 
$x$ value and that this value 
 depends  on $\tau$. Ideally the minimum of $M$ should coincide
with the position of the DHT  at  $t=0$, however this would not be possible
if   in the definition of $M$
only positive times were considered, {\it i.e.} if the limits of the integration were $(0, \tau)$
the dashed trajectory in figure \ref{fig:1}a) would have a lower projection in positive times than the particular solution. An analogous problem would be encountered were only negative times considered, 
that is if the limits of the integration 
would have been $(-\tau,0)$. To determine precisely the position of the DHT at  $t=0$, both positive and negative times must be considered in the definition of $M$. 
Figure 1b) confirms that with this choice the dashed trajectory cannot be distinguished, as it increases its projection
in negative times.
 Figure \ref{fig:2}a) displays   the 
function $M(x^*)_{t=0,\tau}$  evaluated along the trajectories 
(\ref{sol1}), for several $\tau$ values. Figure \ref{fig:2}b) displays the position of the minimum 
of the  function $M_{t=0,\tau}$ 
as a function of $\tau$. These minima correspond to the positions of the $\tau$-distinguished 
trajectories at $t=0$ and as $\tau$ increases
they  approximate the coordinate of the DHT at this time, which is at $x^*=-1$. 
 The pair $(t^l, {\bf x}^l)$ formed by the time 
at which $M$ is computed and the
value of the coordinate ${\bf x}^l$ to which 
the minimum of the  function $M_{t^l,\tau}$ converges for increasing $\tau$  
is called the {\it limit coordinates}.
The graphic \ref{fig:2}b)  illustrates the idea of approaching a point $(t_0, {\bf x}_0)$ of the distinguished trajectory by means 
of the {\it limit coordinates}. In practice the convergence to the {\it limit coordinates} cannot be examined in the limit $\tau  \to \infty$,
either because it is impracticable in a numerical implementation,  
or because in the large $\tau$ limit   errors accumulate, or simply 
because the dynamical system is defined by a finite time data set. For these reasons
the convergence to the {\it limit coordinates} will be tested up to a finite $\tau$. 

\begin{figure}[htb]
\begin{center}
 \includegraphics[width=1.3\columnwidth]{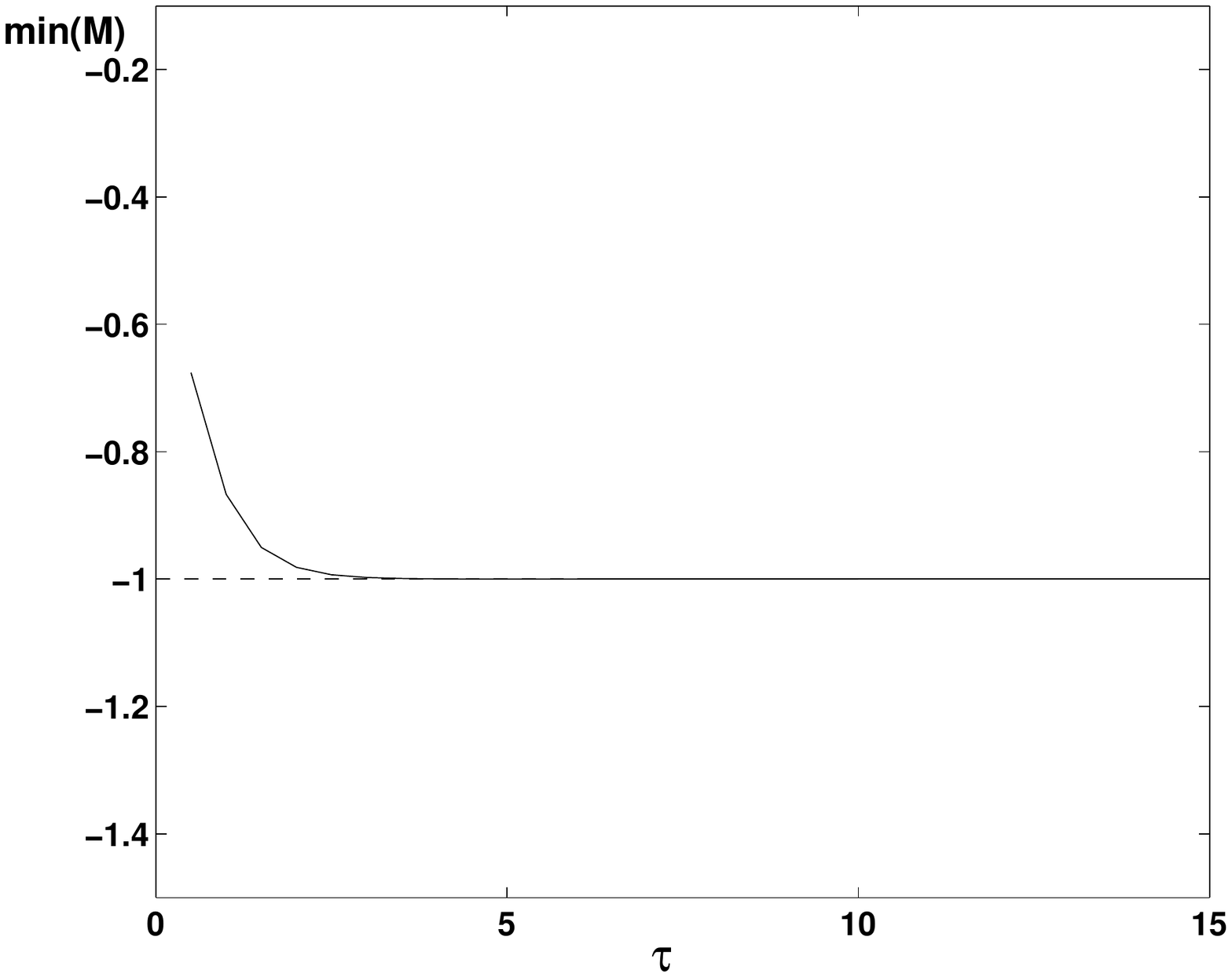}
\end{center}
\vspace{-5cm}
\caption{ }\label{fig:3}
\end{figure}

Figure \ref{fig:2}b)  raises 
the question: what controls the 
rate of the convergence of the minima of $M$ to the coordinates of the DHT?
It is hard to  answer  this question rigorously for a  vector field as general as in Eq. (\ref{adv1}).
However, some insight may be provided by particular examples.
 For instance  the system
\begin{eqnarray}
\frac{dx}{dt}&=&-2x+2t-1,
\end{eqnarray}
has the same DHT as (\ref{eq1b}). Its solution through the point $x^*$ at $t=0$ is  given by
\begin{eqnarray}
x(t)=t-1+e^{-2t} (x^*+1).\label{sol2}
\end{eqnarray}
Here the decay of the solution towards the DHT is faster due to the presence of the exponential 
term $e^{-2t}$. Figure  \ref{fig:3} shows that in this case  the 
rate of the convergence of  the minima of $M$ towards the coordinates of the  DHT 
at time $t=0$ is also faster than before. However
there exist systems in which  the exponential decay of the solution is not a determining
 factor affecting the  rate of the  convergence of the minima of $M$
to the coordinates of the  DHT. For instance, in autonomous systems fixed points are the DTs, and clearly
they are minimizers of $M$ for any $\tau>0$ whatever is  the
exponential rate of growth or decay    of the nearby solution.

In these examples the function $M$ has a unique minimum, but as we will see the situation
will  not always be 
so simple when nonlinearities are involved in the vector field. Also it is important to notice that 
the function $M$ obtained at different $\tau$ values has been  used to obtain the 
{\it limit coordinates} $(t^l_{0},{\bf x}^l_0)$ and that these approach the ${\bf x}_{0}$ coordinate of DHT {\it at a given 
time $t_0$} (here $t_0=t_0^l)$. Once this is obtained,  approaching the DHT at later times $t_k=t_0+ k \Delta t$ would require
applying the same procedure to get the {\it limit coordinates} $(t^l_{k},{\bf x}^l_k)$.
We remark here that the proposed algorithm does not ensure that the set of  {\it limit coordinates}  $(t^l_k, {\bf x}^l_k)$ 
are in fact part of a trajectory. Later we will see that in practice, in many examples these 
points approach a true trajectory,  however in realistic aperiodic flows this has to be 
verified {\it a posteriori}. These considerations lead us to the definition of a {\it Distinguished Trajectory}.
\begin{definition} (Distinguished trajectory). A trajectory $\gamma(t)$ is said to be Distinguished with accuracy $\epsilon$ ($\,0\leq \epsilon$ )
in a time interval $[t_0, t_N]$ if there exists  a continuous path  of {\it limit coordinates} $(t^l,{\bf x}^l)$  where
$t^l \in [t_0, t_N]$, such that,
\begin{eqnarray}
||\gamma(t^l)-{\bf x}^l(t^l)|| \leq \epsilon, \,\, \forall t^l \in [t_0, t_N].
\end{eqnarray}\label{def:dtdef}
Here $|| \cdot ||$ represents the distance defined by
\[||{\bf a}-{\bf b} ||=\sqrt{\sum_{i=1}^{n} (a_i-b_i)^2} \,\,\,\, \,{ with} \,\,\,\,{\bf a}, {\bf b} \in \mathbb{R}^n.\]
\end{definition}
In the numerical exploration of this definition 
we will replace the continuous path  of {\it limit coordinates} $(t^l,{\bf x}^l)$ and the continuous trajectory $\gamma(t)$ by 
discrete representations  $(t^l_{k},{\bf x}^l_k)$ and $\gamma(t^l_k)$ where $t_0\leq t^l_k \leq t_N$.
By definition \ref{def:dtdef}  any trajectory is 
distinguished for sufficiently large  $\epsilon$, however the interesting distinguished trajectories
 are those for which $\epsilon$ is close to zero, which means
it is of the order of the accuracy in which $\gamma(t^l_k)$ and ${\bf x}^l(t^l_k)$ are numerically determined,
or zero, if an exact expression is known for both. 

Underlying definitions 2 and 3 is the geometrical idea that distinguished trajectories, 
which act as organising centres of the flow in phase space, are those 
that "move less" (in a certain sense) than other nearby trajectories.
 This property of ``moving less'' is 
satisfied by minima of the  function $M$ as it measures 
the length of the displacement in phase space of a trajectory forwards and backwards in time.  
In fact this property is related somehow to property (2) of the definition 
provided in \cite{kayo} and presented at the beginning of Section 2, as the trajectory that ``moves least'' 
is not  expected to leave  the neighbourhood $\mathcal{B}$.

Definitions \ref{def:dt} and \ref{def:dtdef} are made for a general dynamical system in any dimension $n$. 
The purpose of this paper is the exploration of these definitions, but more in an illustrative than demonstrative way, 
as it is impossible to provide examples 
for every possible $n$, and  one cannot deal with every possible example at a given $n$. 
Even if one wants to provide a rigorous formal proof that  the definition  recovers specific trajectories
such as periodic orbits (it is not obvious that in general they have to satisfy our definition), 
this has to be done with some further hypotheses 
on the vector field and  proofs
will not be valid beyond the assumed hypotheses. Therefore we restrict the
 discussion to dimensions up to 3, as these are the  dimensions important for 
geophysical flows, which are what originally  
motivated the definition. However  it is sensible to make  the same definition 
for any  dimension $n$, as it is clear that it works
for autonomous systems of any dimension.
Fixed points are  the kind of trajectory  expected to be recovered 
by the definition and they do not move at all in the phase space.
For these  $M=0$, while $M>0$ for any other trajectory in the neighbourhood which is not a fixed point.


We conclude this section with some remarks. First, it is not guaranteed {\it a priori} that for an arbitrary
vector field, satisfying only some  rather general hypotheses such as those of Eq. (\ref{adv1}), 
the function $M$ will have  a minimum, however this is not a problem from the point of view
of  the definition.
For instance the same thing happens for general non-linear autonomous systems.
 In these systems fixed points are perfectly
defined although one does not know {\it a priori} if such points  exist for arbitrary examples.
If they exist, it is possible to find them by either solving the nonlinear equation ${\bf v}({\bf x})=0$, 
or by applying  definitions
\ref{def:dt} and \ref{def:dtdef}.
In the same way one does not know  {\it a priori} if distinguished trajectories  exist for 
a general vector field
however if they exist they can been found with the tools proposed in this article.  
Second, even if a path of {\it limit coordinates} is found it 
is not guaranteed that it will be a trajectory, although if that is the case then 
from definition \ref{def:dtdef} follows  that this trajectory is distinguished.
Third, one might think that if {\it limit coordinates} are found 
at $t_0$  that  approach with great accuracy a point of
{\it an existing DT}, then the iterative procedure described above for finding 
a set  of {\it limit coordinates} $(t^l_{k},{\bf x}^l_k)$
approaching the DT at later times is an unnecessary computational effort, 
as those coordinates could have been equally well obtained 
by integrating forwards the initial data. However there exist examples such as  a hyperbolic  
DT in dimension greater than one with at least 1D unstable manifold, that cannot be integrated like this, as 
 the integrated trajectory will
eventually leave the neighbourhood of the DT through the unstable manifold no matter how small the initial error is. 
In summary  the proposed methodology
based on {\it limit coordinates} provides a systematic way of finding DT, which 
can be elusive and difficult to obtain. 
We will discuss these issues in detail in later sections.


\section{A numerical algorithm}

In this section we propose an algorithm for computing 
a path of {\it limit coordinates} in a time interval, and 
we verify that it is close to a DT of a known example.
For this purpose  we calculate, at increasing $\tau$ values, the minimum of the function $M_{t=0,\tau}({\bf x})$ for ${\bf x}$ in  an open set in $\mathbb{R}^n$. 
The method is  illustrated in a  2D case, the periodically forced Duffing equation
\begin{eqnarray}
\dot{x}&=&y,\nonumber \\
\dot{y}&=&x-x^3+\varepsilon\sin (t),\label{eq:duffing}
\end{eqnarray} 
where $\varepsilon$ is a small parameter. The hyperbolic fixed point of the unperturbed  autonomous system (i.e., $\varepsilon=0$)
is at the origin ${\bf x}= (0,0)$. For small $\varepsilon$, it is 
possible to compute by perturbation theory (see \cite{physd}), 
the following periodic trajectory which stays close to the origin: 

\begin{equation}\label{eq:dhtduffing}
{\bf x}_{DHT}(t)=-\frac{\varepsilon}{2} \binom{\sin t}{\cos t}-\frac{\varepsilon^3}{40}
\binom{2\sin^3 t+\frac{3}{2}\sin t\cos^2 t}{\frac{3}{2}\cos^3 t+3\sin^2t\cos t}
+\mathcal{O}(\varepsilon^5).
\end{equation}
For $\varepsilon=0.1$,  Eq. (\ref{eq:dhtduffing})  
is accurate   up to the fifth digit. This trajectory  is  identified as {\it Distinguished} in Ref. \cite{kayo}, for this reason we 
have labelled it a $DHT$.  
Substituting the expression,
\begin{equation}
{\bf x}=(x,y)={\bf x}_{DHT}(t)+\left(\xi_1,\xi_2 \right)
\end{equation}
into Eq. (\ref{eq:duffing}) and by dropping the nonlinear terms one finds that the linearized
equations have two linearly independent solutions in terms of  which the time
evolution of the components ${(\xi_1, \xi_2)}$ is: 
\begin{equation}
\left(\xi_1,\xi_2 \right) = \alpha \,{e}^{t} \left(\begin{array}{c}1/\sqrt{2}\\ 1/\sqrt{2} \end{array}\right)+\beta \, {e}^{-t} \left(\begin{array}{r}-1/\sqrt{2}\\ 1/\sqrt{2}\end{array}\right)+\mathcal{O}(\varepsilon^2).\label{timeev}
\end{equation}
Eq. (\ref{timeev}) confirms the hyperbolicity of the solution (\ref{eq:dhtduffing}).

\begin{figure}[htb]
\begin{center}
\includegraphics[width=0.8\columnwidth]{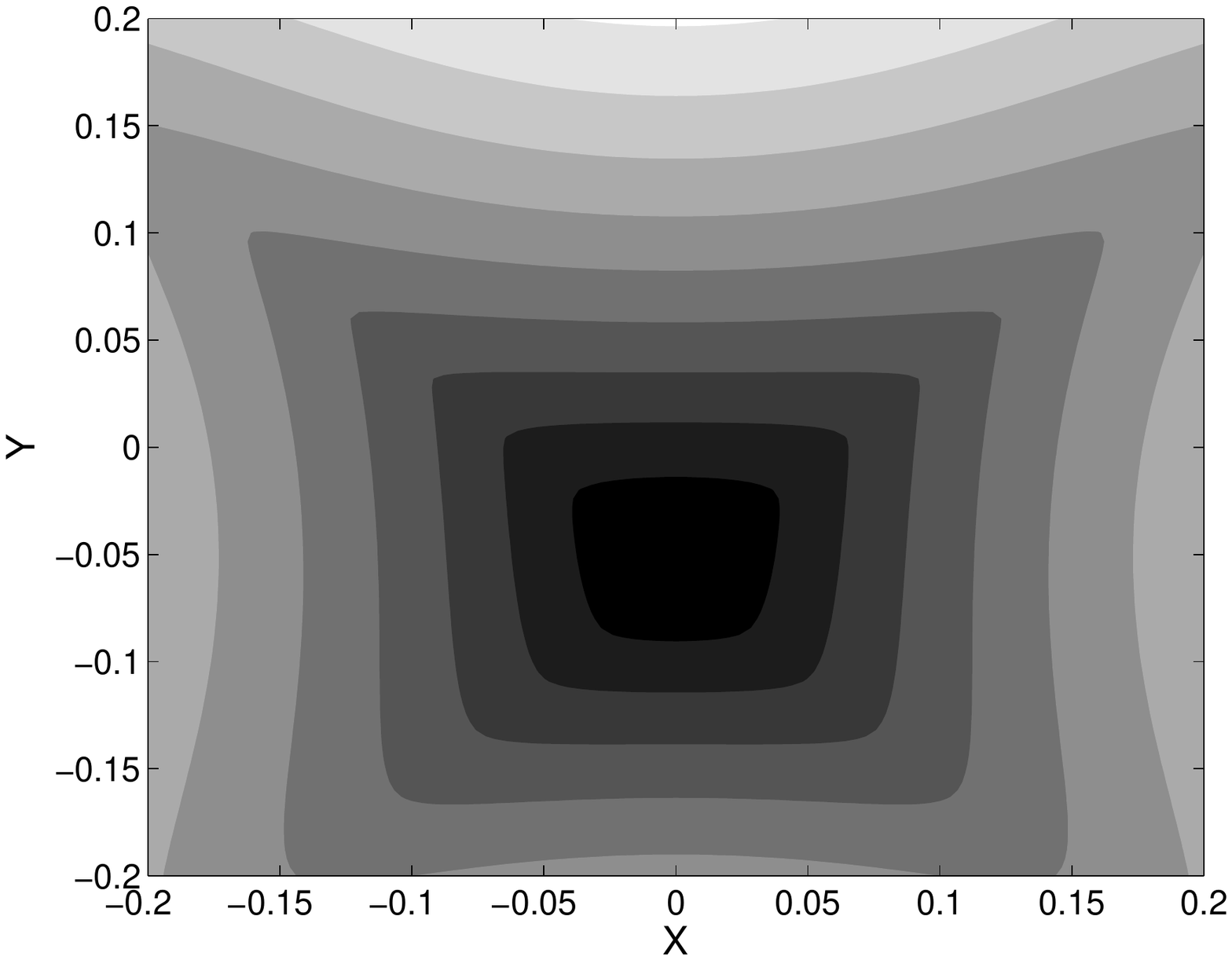}
\end{center}
\caption{}\label{fig:duffinggrid}
\end{figure}

This explicit expression for the distinguished hyperbolic trajectory is
a benchmark for testing the utility of our definition. 
The procedure starts by determining the coordinates of ${\bf x}_{DHT}$ at time $t=0$.
We consider the open set ${\mathcal D}\subset \mathbb{R}^2$, defined by
 ${\mathcal D}=(-0.2,0.2)\times(-0.2,0.2)$ and in the function $M_{t=0,\tau}({\bf x})$ we take $\tau$ 
to be $2$. 
Figure \ref{fig:duffinggrid} displays a contour plot of $M_{t=0,\tau=2}({\bf x})$ which has a minimum  at ${\bf x}=(0, -5.7057  \cdot 10^{-2})$. $M_{t=0,\tau=2}({\bf x})$ quantifies displacements of particles in  phase space, and its minimum corresponds to 
the initial condition that 
``moves less'' over the $\tau$ interval $[-2, 2]$.
As noted in the previous section,  when the
 value of $\tau$ is increased,   
the position of the minimum gets closer and closer to  the coordinates of the DHT at $t=0$. 
 Figure \ref{fig:duffingtau} shows contour plots of the function $M$ for several $\tau$ values.
Fig.  \ref{fig:duffingtau}a) displays a typical hyperbolic structure for $M$
 for $\tau=5$ where the directions of the stable and unstable manifolds are easily recognised.
In Fig.  \ref{fig:duffingtau}a) the function $M$ has a unique minimum  at  ${\bf x}=(0, -4.979 \cdot 10^{-2})$ 
while in Fig.  \ref{fig:duffingtau}b) there  appear several minima for $\tau=10$. 
The global minimum in this picture corresponds to ${\bf x}=(0, -5.0042565261 \cdot 10^{-2})$.
Figure \ref{fig:duffingtray}a)  compares the $x$-coordinate of ${\bf x}_{DHT}$
as a function of time with  trajectories having initial conditions
at the global minima of  $M_{t=0, \tau=2}$ and $M_{t=0, \tau=10}$.
Taking  as initial condition the global minimum of $M_{t=0, \tau}$  for $\tau=10$  
provides a trajectory that  stays  close to
 ${\bf x}_{DHT}$ for a longer time interval than  for $\tau=2$, which
confirms  that larger $\tau$-values more closely approach the coordinates of the DHT. 
Fig.  \ref{fig:duffingtau}c) displays the contour plot of $M_{t=0,\tau=50}({\bf x})$. Its 
global minimum is  at ${\bf x}=(0,-5.0037606418 \cdot 10^{-2})$. 
The associated trajectory depicted in Fig. \ref{fig:duffingtray}b) shows that 
this initial condition tracks the DHT for a longer time interval than those obtained for $\tau=2$ and 10,
however the figure shows that  the integration of the DHT 
in  $(-50, 50)$ is not possible. In fact the associated 
trajectory stays close to the DHT  only in  the time interval $(-20, 20)$. This confirms that results
obtained for $\tau=50$ are the same as those obtained for $\tau=20$. 
In practice for a finite precision  numerical scheme, 
such as 
a 5th order Runge Kutta used here, the approach to the DHT has  
an upper bound depending on $\tau$. This occurs because the stable and unstable manifolds of the  hyperbolic 
trajectory magnify any initial error in either negative  or positive  time and 
beyond this $\tau$-limit numerical  errors dominate. The convergence towards 
the DHT is confirmed in Fig. \ref{fig:duffingxy} which
displays the evolution of the coordinates $x$ and $y$ of the global minimum of $M$ as a function of the parameter $\tau$. 

New minima appearing in Figs.  \ref{fig:duffingtau}b) and c)
relate to the existence of different $\tau$-distinguished trajectories. 
As illustrated in Fig. \ref{fig:duffingtray}b), they correspond to trajectories which stay close to
${\bf x}_{DHT}$ in a small time range contained in the interval $-\tau < t < \tau$, 
but which later fly apart from the DHT. 

We now describe a numerical scheme to compute a path of limit coordinates. 
The algorithm has the following steps:

\begin{enumerate}
\item Step 1. Discretize the domain  ${\mathcal D}$ at the initial time $t=t_0$ at which
one wishes to compute a DT. For instance, the grid size of this domain in Fig.  \ref{fig:duffinggrid}
is $101 \times 101 $. The function $M$ is evaluated  at each grid point for a given  $\tau_0$.
\item Step 2. Search for the local minima of  $M_{t_0,\tau_0}$ in the interior of the
grid. These minima approach   the coordinates of $\tau_0$-distinguished trajectories within the accuracy of the grid.
In what follows we restrict our description to the case of a unique minimum, 
as this simplifies the description;
the procedure is easily generalised to the case of multiple minima. 
\item  Step 3. Improve the approach of the coordinates of the $\tau_0$-distinguished trajectory up to precision $\delta$. 
For this purpose build up a
 $3^n$ grid 
centred on the candidate point provided by step 2, (for the 2D case this is a $3\times 3$ grid as Fig. \ref{fig:epsilongrid} illustrates), 
setting the  distance between nodes equal to $\delta$.
Then evaluate $M_{t_0,\tau_0}$ at the points of the $\delta$-grid. If the minimum  of $M_{t_0,\tau_0}$ is in the interior
of the grid, then the coordinates of the $\tau_0$-distinguished trajectory are known to within $\delta$ accuracy. Otherwise
the $\delta$-grid must be rebuilt centred on the boundary point where the minimum has been located, and
$M_{t_0,\tau_0}$ must be re-evaluated in the new $\delta$-grid. This procedure stops when the minimum  
of $M_{t_0,\tau_0}$ is in the interior of the grid.
\item  Step 4. Computing the {\it limit coordinates} at time $t_0$.
Define a sequence of increasing $\tau$-values as follows: $\tau_1=\tau_0+ \Delta \tau$ and $\tau_2=\tau_0+ 2 \Delta \tau$.
Then evaluate $M_{t_0,\tau_0}$, $M_{t_0,\tau_1}$ and $M_{t_0,\tau_2}$
on the $\delta$-grid. If  the minimum is at  an interior position for the three cases,
then we consider that {\it limit coordinates} have been found within  $\delta$  accuracy.
We note that this is a necessary but not sufficient condition as one does not know {\it a priori}
the convergence  rate to the distinguished trajectory. Although this criterion could be strengthened,
it has been tested and found to be adequated for the examples explained in subsequent sections.
If the condition  defined above of having a minimum  at  an interior position for the sequence of $\tau$-values
is not satisfied then, after  replacing $\tau_0$ by $\tau_1$, we return to step 3 and then to step 4.
The loop between steps 3 and 4 is stopped when the condition of step 4 is satisfied for some $\tau_k$.
\item Step 5. Compute the {\it limit coordinates} at time $t_1=t_0 + \Delta t$. Once the {\it limit coordinates}
have been approached
at time $t_0$, they are  integrated forward  numerically up to time $t_1$. If the {\it limit coordinates} converge to a  
hyperbolic  DT 
with an unstable manifold, the position ${\bf x}(t_1)$ obtained should  deviate from the position 
of the DT at time $t_1$. In order to correct this, the procedure described above is repeated from step 3 onwards.
For that purpose in the definition of $M$, $t_0$ is replaced by $t_1$ and the $\tau$-value 
is reset to $\tau_0$. The coordinates ${\bf x}(t_1)$ are the first
 approximation to the  $\tau_0$-distinguished trajectory at time $t_1$. Once the {\it limit coordinates}
are found
for time $t_1$ it is possible to repeat the procedure  to locate them at successive times 
 $t_2,t_3,\ldots,t_N$.
\end{enumerate}

The  algorithm  requires as inputs: an explicit expression for
the dynamical system (\ref{adv1}); the definition of the domain $\mathcal{D} \subset \mathbb{R}^n$; the initial and final
times $t_0$, $t_N$ at which DTs are required, 
and the time step $\Delta t$ for intermediate times; the initial $\tau_0$
and the increment $\Delta \tau$; the precision $\delta$. As an output the algorithm gives
a path of limit coordinates  at the selected times $t_k$.

Next we discuss in more detail some technical aspects related to the implementation
of the above algorithm. Steps 1 and 3 require evaluating 
$M_{t_0,\tau_0}$ as defined in Eq. (\ref{def:M}). We explain how this is done for the contour plots displayed 
in Figs.  \ref{fig:duffinggrid} and \ref{fig:duffingtau}, which  refer 
to the system (\ref{eq:duffing}) at $t_0=0$.
Fig. \ref{tra1} shows a schematic projection onto the $\mathbb{R}^2$ plane of a possible trajectory
${\bf x}(t)$ of the system  from $-\tau$ to $\tau$. 
As it was obtained  numerically, only a finite number of points ($L$) appear. This picture suggests the following  discrete version of Eq. (\ref{def:M}) for $M$:
\begin{equation}
M({\bf x})_{0,\tau}= \sum_{j=1}^{L-1} \left(  \int^{p_f}_{p_i} \! \!\!\sqrt{\left(\frac{d x_j(p)}{dp}\right)^2+\left(\frac{d y_j(p)}{dp}\right)^2 }dp \right), 
\end{equation}
where the functions $x_j(p)$ and $y_j(p)$ represent a curve interpolation parametrized by $p$, and the integral
\begin{equation}
\int^{p_f}_{p_i} \! \!\!\sqrt{\left(\frac{d x_j(p)}{dp}\right)^2+\left(\frac{d y_j(p)}{dp}\right)^2 }dp
\end{equation}
is computed  numerically. In our case we use the Romberges method (see \cite{nr}) of  order $2K$ with $K=5$. 
It is clear that the accuracy of the evaluation of $M$ will depend on the number of points 
 on the trajectory  $L$, which is controlled by the size of the time step, $h$, of the integrator (a 5th order Runge Kutta method)
and on the interpolation scheme between points. 
Two interpolation methods are compared in tables (\ref{tab:erroresl}) and (\ref{tab:erroresd}).
Results in table (\ref{tab:erroresl}) are obtained with linear interpolation between nodes. Results in  
table (\ref{tab:erroresd}) correspond to the interpolation method used by Dritschel \cite{dr} in the context of contour dynamics,
 which has been successfully applied in \cite{physd} to the computation of invariant manifolds 
for aperiodic flows. 
This method interpolates a piece of the curve in Fig. \ref{tra1} between consecutive nodes as follows: 
\begin{equation}
{\bf x}_j (p) = {\bf x}_j + p {\bf t}_j + \eta_j (p){\bf n}_j \label{eq:intdri}
\end{equation}
for $p_i=0 \leq  p \leq  p_f=1$ with ${\bf x}_j (0) = {\bf x}_j$ and ${\bf x}_j (1) = {\bf x}_{j+1}$, where:
\begin{eqnarray}
{\bf t}_j &=& (a_j , b_j ) = {\bf x}_{j+1} - {\bf x}_{j},  \,\,\, {\bf t}_j  \in \mathbb{R}^2\\
{\bf n}_j &=& (-b_j , a_j ),   \,\,\,  {\bf n}_j \in  \mathbb{R}^2\\
\eta_j (p) &=& \mu_j p + \beta_j p^2 + \gamma_j p^3, \,\,\,  \eta_j \in    \mathbb{R}.
\end{eqnarray}
The cubic interpolation coefficients $\mu_j$, $\beta_j$ and $\gamma_j$ are
\[
 \mu_j = - \frac{1}{3} d_j \kappa_j - \frac{1}{6} d_j \kappa_{j+1}, \,\,\, \beta_j = 2 d_j \kappa_j, \,\,\, \gamma_j = \frac{1}{6}d_j 
(\kappa_{j+1} - \kappa_j ),
\]
where $d_j = |{\bf x}_{j+1} - {\bf x}_j |$ and
\[             
\kappa_j = 2 \frac{a_{j-1} b_j - b_{j-1} a_j}{|d^2_{j-1} {\bf t}_j + d^2_j {\bf t}_j-1 |}                                         \]                   
is the local curvature defined by the circle through the three points, $x_{j-1}$, $x_{j}$, and $x_{j+1}$.

Tables (\ref{tab:erroresl}) and (\ref{tab:erroresd}) show the errors in 
the computed lengths of the ellipses for different  ratios of
major to minor axis.  The reference length is that obtained with GNU Octave 
version 2.1.73, as it provides 16 correct digits for the known circumference. 
The tables confirm that the Dritschel's method is superior to linear interpolation and 
it is the one used to
compute the function $M$. In the trajectory from $-\tau$ to $\tau$ the number of points $L$ is determined by the time step size 
of the Runge Kutta method which is set to $10^{-2}$.

Another important element  of the algorithm needing
 discussion is the value of the input parameters, in particular of $\tau_0$ and
$\Delta \tau$. It is clear from Fig. \ref{fig:duffingtau} that large $\tau$ values are not convenient as they
increase the roughness of the function 
$M$ and  several local minima may appear in the neighbourhood of a DHT  that correspond 
to trajectories that stay close to it
for some time. On the other hand it is clear that sufficiently large  $\tau$ values are required to fix the coordinates of the DHT to within prescribed accuracy. Combining these observations suggests the 
use of relatively small values for  
the initial $\tau_0$.
In the example above $\tau_0=2$, provides, as a starting point, a smooth $M$  as that of Fig. \ref{fig:duffinggrid}. 
The increments should not be large. In practice we have chosen $\Delta \tau=\tau_0/2$. This prevents 
from stepping to a  too rough $M$ before getting close enough to the sought after DHT. Some of the  local
minima appearing in Fig. \ref{fig:duffingtau}b) are just apparent and disappear with a more refined grid. 
However, as already observed,
others  belong to  true $\tau$-distinguished trajectories, which are secondary and can be  avoided
if  the increment of the $\tau$-values is conveniently small. These choices are found to be appropriate 
for determining with great accuracy the DHT in (\ref{eq:dhtduffing}) by means of a path of {\it limit coordinates}.  
Figure \ref{fig:dhtduffing}a) represents both the 
analytical DHT and the numerical limit coordinates 
and Figure \ref{fig:dhtduffing}b) displays the distance  between the 
exact and the numerical approach, confirming that the DHT in (\ref{eq:dhtduffing}) is also a DT 
in the sense of our definition 
\ref{def:dtdef} with
accuracy $\epsilon=10^{-6}$. Other parameters in the
algorithm are: $\delta=10^{-6}$, step size in the Runge Kutta method $ h=10^{-2}$,  $t_0=0$, $t_N= 6$, and $\Delta t=0.01$.
To locate the DHT with accuracy $\delta=10^{-6}$ requires increasing values of $\tau$  up to  15, which is
near  the limit of the integration method. Figure \ref{fig:dufftau}
shows the maximum $\tau$ required at each $t_k$.

\section{Applications to exact examples}

In this section we apply the algorithm explained in the previous section to selected examples.

\subsection{ A non-hyperbolic distinguished trajectory}

The unperturbed autonomous system (\ref{eq:duffing}) obtained with $\varepsilon=0$ has  non-hyperbolic
fixed points at $(-1,0)$ and $(1,0)$. Obviously these fixed points correspond to DTs 
which are also $\tau$-distinguished trajectories for all $\tau>0$. 
For the periodically forced system (\ref{eq:duffing}) with
small $\varepsilon$ it is possible  using perturbation theory
to find periodic solutions close to these fixed points in a manner similar to the analysis
of the hyperbolic example  made in the previous section.
For instance close to the point $(1,0)$ we find the periodic trajectory:
\begin{equation}\label{eq:detduffing}
x_{DET}(t)=-\binom{1}{0}+\varepsilon \binom{\sin t}{\cos t}+3\varepsilon^2
\binom{\frac{1}{2}\cos^2 t}{-\sin t\cos t}
+\mathcal{O}(\varepsilon^3).
\end{equation}
This solution has not been considered {\it Distinguished} in previous works \cite{kayo,yu}, as these have been focused
on hyperbolic trajectories and this solution, as is proved next,  is not hyperbolic. 
However, in anticipation of its having the {\it Distinguished} 
property, we have labelled it $DET$ for two reasons.
 One is that it is periodic, and we expect
periodic orbits to be distinguished, and second is that it is in clear correspondence to the elliptic fixed point (-1,0)
in the case $\varepsilon=0$, and fixed points are DTs.

To determine the  stability of  (\ref{eq:detduffing}) we proceed as before, by substituting the expression
\begin{equation}
{\bf x}=(x,y)={\bf x}_{DET}(t)+\left(\xi_1,\xi_2 \right)
\end{equation}
into Eq. (\ref{eq:duffing}). We find that the  linearized system at order $\varepsilon^0$ is:
\begin{eqnarray}
\frac{d \xi_1}{d t}&=&\xi_2\\
\frac{d \xi_2}{d t}&=&-2\xi_1.
\end{eqnarray}
Therefore the linearized flow around ${\bf x}_{DET}(t)$ evolves according to:
\begin{equation}
\left(\xi_1,\xi_2 \right) = \alpha \,{e}^{i\sqrt{2}t} \left(\begin{array}{c}1/\sqrt{3}\\ i\sqrt{2/3} \end{array}\right)+\alpha^* \, {e}^{-i\sqrt{2}t} \left(\begin{array}{r}1/\sqrt{3}\\ -i\sqrt{2/3}\end{array}\right)+\mathcal{O}(\varepsilon),
\end{equation}
which clearly is not hyperbolic. Here $\alpha$ and $\alpha^*$ are complex conjugate numbers.

We apply our algorithm to determine the {\it limit coordinates} 
approaching (\ref{eq:detduffing}), as we want to verify whether 
 definition~\ref{def:dtdef}
also  works for time-dependent non-hyperbolic solutions. The following input
 is considered: ${\mathcal D}=(-1.2,-0.8)\times(-0.2,0.2)$, $\tau_0=2$, $\Delta \tau=1$, $\delta=10^{-4}$,
$t_0=0$, $t_N=6$, and time step $10^{-2}$ for the Runge-Kutta integrator. 
We note that the accuracy $\delta$ is not
as demanding as before, since now the exact ${\bf x}_{DET}$ for $\varepsilon=0.1$
is only accurate up to the third digit.
Figure \ref{fig:duffingetau} shows a rather different structure for the function $M$. An important 
feature is the smoothness of $M$ close to the DET even for large $\tau$. 
In figure \ref{fig:duffingetau}b)  the differences 
between the  rather flat region around the position of the DT given by (\ref{eq:detduffing}), which appears in the dark tone,
and the roughness of the outer part are remarkable. The irregularity of this region suggests that inside it nearby trajectories follow
rather different paths as  happens for chaotic motions, while the regularity of the central core suggests the 
existence of trapped trajectories circling around  the  DET. From 
this perspective the function $M$ for large $\tau$
seems a useful tool for fixing 
the boundaries of a Lagrangian eddy, different to the methods proposed in \cite{br,haller3}.

Figure \ref{fig:duffingexy} shows the rate of convergence to the global minimum of $M$  in the domain  ${\mathcal D}$
as a function of $\tau$. The convergence towards the coordinates of the DT  is oscillatory and rather 
slow since $\tau$ values up to $600$ are required. A slight difference between the exact coordinates of the DT and
the numerically computed limit coordinates is evident, however we note that these differences are 
consistent with  the precision to which the exact DT is known, which is only  to the third digit.
Figure  \ref{fig:detduffing}, and more specifically  figure  \ref{fig:detduffing}b),
confirms that the exact expression in Eq. (\ref{eq:detduffing})
is in fact a distinguished trajectory according to our definition~\ref{def:dtdef} with 
accuracy $\epsilon=4 \cdot 10^{-3}$.
 
Figure \ref{fig:duffingetray}a) shows a forward and backward integration along the time interval $(-50, 50)$
taking as  initial data the limit coordinates supplied by our algorithm 
at time $t_0=0$, and compares it with the exact solution of the DT.
From figure \ref{fig:duffingetray}b) it can be seen that this trajectory evolves  close to the exact 
solution in the entire time range.
This result shows that contrary to what happens near hyperbolic trajectories,
near non hyperbolic trajectories,  small error does not amplify and as consequence,
once a DT {\it is known to exist} it could have been computed simply by integrating forwards and backwards
the limit coordinates found at a given time $t_k$. However one needs to be careful here, as a trajectory
is not necessarily distinguished at all times, and for it to be properly called
distinguished, it  should be verified that it stays close to the
{\it limit coordinates} in the whole time interval, and therefore one cannot avoid computing {\it limit coordinates}
along the time interval in this case either. We will return to this point in the next section.
   
\subsection{ The rotating Duffing equation}

Next we analyse the aperiodic hyperbolic distinguished 
trajectory of a system already studied in \cite{physd},  the rotating Duffing equation:
\begin{eqnarray}
\left(\begin{array}{c}\dot{\eta_1}\\ \dot{\eta_2} \end{array}\right) &=& \left(\begin{array}{cc}\sin 2 \omega t & \cos 2\omega t +\omega\\
\cos 2\omega t -\omega & -\sin 2 \omega t
 \end{array}\right)\left(\begin{array}{c}\eta_1\\ \eta_2 \end{array}\right) \nonumber\\
&+&( \varepsilon \sin t - [ \cos \omega t 
\eta_1 - \sin \omega t \eta_2 ]^3 )\left(\begin{array}{c}\sin \omega t \\ \cos \omega t \end{array}\right).
\end{eqnarray}
This Duffing equation is quasi-periodic in time when the rotation rate $\omega$ is irrational. It is obtained from 
the system (\ref{eq:duffing}) by applying the rotation  ${\bf x} = R(t) \boldsymbol{\eta} $, where
\begin{eqnarray}
R(t)=\left(\begin{array}{cc}\cos \omega t & -\sin \omega t \\
\sin \omega t  & \cos \omega t
 \end{array}\right).
\end{eqnarray}
The DHT can also be obtained through the coordinate transformation:
\begin{equation}
{\boldsymbol{\eta}}_{DHT} (t) = R(t)^{-1} {\bf x}_{DHT} (t). \label{dhtrduff}
\end{equation}
Figure \ref{fig:dhtrduffing}, and in particular Fig. \ref{fig:dhtrduffing}b), confirms that the DHT (\ref{dhtrduff})
is also a DHT according to our definition ~\ref{def:dtdef} with accuracy $\epsilon=4 \cdot 10^{-6}$.

\subsection{ A 3D extension of the Duffing equation}

In this section we apply our definitions to an example in higher dimension. In particular we
consider a 3D extension of the Duffing equation:
\begin{eqnarray}
\dot{x}&=&y,\nonumber \\
\dot{y}&=&x-x^3+\varepsilon\sin (t),\label{eq:duffing3d}\\
\dot{z}&=&z+\varepsilon\sin (t).\nonumber
\end{eqnarray} 
The hyperbolic fixed point of the unperturbed  autonomous system (i.e., $\varepsilon=0$)
is at the origin ${\bf x}= (0,0,0)$. The solution   for small $\varepsilon$ becomes: 
\begin{eqnarray}
{\bf x}_{DHT}(t)&=&-\frac{\varepsilon}{2} \left(\begin{array}{c} \sin t\\ \cos t \\ {\small \cos t - \sin t}\end{array}\right)\nonumber\\&-&\frac{\varepsilon^3}{40}
 \left(\begin{array}{c} {2\sin^3 t+\frac{3}{2}\sin t\cos^2 t}\\{\frac{3}{2}\cos^3 t+3\sin^2t\cos t}\\ 0\end{array}\right)\label{eq:dhtduffing3d}
+\mathcal{O}(\varepsilon^5).
\end{eqnarray}

The numerical scheme explained in Section 3 is easily adapted to higher dimensions. 
However some changes must be made.
The computation of $M$ requires approximating lengths of  trajectories which in 3D  
needs an  interpolation scheme different to that 
of Eq. (\ref{eq:intdri}), which is only valid in $\mathbb{R}^2$. We consider the linear interpolation instead. 
This interpolation evaluates the function $M$ satisfactorily if trajectories are represented by  a large number of points. 
This is achieved by using a Runge-Kutta  method with time step  $h=10^{-4}$.
Figure \ref{fig:duffingtau3d} indicates the evolution of  coordinates associated with the minimum of $M$ 
as a function of $\tau$ (solid line). The dashed line corresponds to the exact perturbative solution. There is evident 
a clear  convergence towards the exact position although there is a significant jump in the asymptotic behaviour
beyond $\tau \sim 50$. This jump  is due to round off errors in the determination of $M$ for large $\tau$. 
The third equation in (\ref{eq:duffing3d}) is just a linear equation and for this reason
 solutions which are in the neighbourhood of the DHT have $z$-coordinate
 growing exponentially  in backwards time. Thus for large $\tau$ values, the evaluation of  $M$ is made along  very long trajectories in the $z$-coordinate, 
which are underrepresented by points   sampled every $h=10^{-4}$ (see table I) and where lengths 
are badly calculated  by adding up  very small and very large (and inaccurate) numbers.
In spite of  this, figure \ref{fig:duffing3d} confirms that  the exact distinguished
trajectory can be accurately obtained with our methodology and that for $\tau<50$ errors are within the expected margin.
The remaining input parameters used  in Figs. \ref{fig:duffingtau3d} and \ref{fig:duffing3d} 
are: ${\mathcal D}=(-0.2,0.2)\times(-0.2,0.2)\times(-0.2,0.2)$, $\tau_0=2$, $\Delta \tau=1$, $\delta=10^{-6}$, 
step size $ h=10^{-4}$ in the Runge Kutta method, $t_0=0$, $t_N= 6$, and $ \Delta t=10^{-2}$.

As we explain next, the computational demands made by this example are considerably larger than 
they were for the previously considered 2D examples. 
When  determining a DT, most of the CPU time is spent computing the value of $M$ on
 the $\delta$-grid  displayed in Fig. \ref{fig:epsilongrid}. The number of neighbours
of the interior point grows with the dimension $n$ as $3^n$, therefore
when  the problem increases its dimension from $n$ to $n+1$, the computational demands are multiplied by 3. 
Another factor that contributes to increased computational time is the decrease 
of the Runge Kutta time step $ h$
in the evaluation of trajectories on the $\delta$-grid. This increases the number of points in the trajectory 
(and therefore the number of  operations) with respect to the previous 
Dristchel approach by a factor $100$. This factor  is partially balanced by the fact that for the same number of points
the arclength is computed more rapidly with the linear than with the Dristchel interpolation.




\section{Application to vector fields defined as finite time data sets}

In this section we explore definition \ref{def:dtdef} for a highly aperiodic 2D  flow in which 
the vector field is defined as a finite time data set. In particular we consider the output of 
a  quasigeostrophic wind-driven double gyre  model in a regime already studied in \cite{physrep,nlpg}.
Details of this model may be found in  \cite{nlpg, coulliete}. Fig.~\ref{fig:sfqgn} shows a typical
output for the streamfunction provided by this model.
The velocity data set is obtained on a 1000 km $\times$ 2000 km rectangular domain and spans 4000 days. This interval is considered for a fluid started from rest and allowed to spin for 25000 days.
Free slip  conditions are considered for the velocities on the boundaries
 and the wind stress curl is 0.32 dyn/cm$^2$.
The equations of motion for this system are given by:
\begin{eqnarray}
\dot{x}&=&v_x(x,y,t)=-\frac{\partial \psi}{\partial y},\\
\dot{y}&=&v_y(x,y,t)=\frac{\partial \psi}{\partial x},
\end{eqnarray}
and the variables $x$ and $y$ are in the rescaled domain $[0, 1]\times [0, 2]$. Here the  velocity fields $v_x$ and $v_y$ are provided as a finite time data set 
and are interpolated using bicubic interpolation in space and 3rd order Lagrange polynomials in time. 
This method
has been reported to be good enough for integrating trajectories in \cite{cf}.
We will focus our analysis in the time interval $[0, 900]$ in the area marked by a rectangle in Fig. \ref{fig:sfqgn}
for which \cite{nlpg} reports the computations of 
several DHTs. In \cite{nlpg}  distinguished trajectories are computed by means of an iterative algorithm 
which is initialized  on a hyperbolic instantaneous stagnation point (ISP). In particular two
 paths of such  ISPs are chosen in the Northern gyre in the time intervals $[0, 339]$ and $[446, 880]$.
From each  of these paths, a DHT is computed which is  
in the same geographical area although its coordinates are determined for a  different time range. 
In Fig. \ref{fig:dhtqgnab} we show the $x$ and $y$ evolutions for these trajectories. These coordinates have been computed 
with a different algorithm  to that proposed in \cite{nlpg}. Instead each corresponds to a trajectory 
which is in the intersection of  a piece of a stable manifold and a piece of an unstable manifold which are 
evolved in  backwards and forwards time respectively. 
In this procedure, in order to avoid the numerous intersections 
between stable and unstable manifolds, which make difficult the tracking of the trajectory which is distinguished,
 manifolds are trimmed at each time step following the ideas  in \cite{physrep} where 
a method is described to compute a piece of single
branch of  the stable or of the unstable manifold. This method  takes advantage of the fact that a
DHT must be in the intersection of both manifolds at all times, as it is a trajectory, however 
does not improve the method
explained in  \cite{nlpg}  in the sense that it does not 
allow either to extend the computation of  the DHT beyond the time interval in which the ISP exists.
Many questions have been raised for these trajectories as has been discussed in \cite{physrep,nlpg}.
For instance as 
they have been computed only in finite time intervals on which the ISP exists, one can ask how to pursue
its computation beyond that interval.
Another open issue in \cite{nlpg} concerns deciding if  the two DHT  in Fig. \ref{fig:dhtqgnab}
computed at different times 
are part of the same  trajectory. 
In \cite{physrep}, the question is raised of whether it can happen that a DHT ceases to be distinguished or hyperbolic. In this section we apply our algorithm to compute limit coordinates
and verify whether trajectories in Fig. \ref{fig:dhtqgnab} are distinguished or not following our definition
\ref{def:dtdef}. Also we will describe how this definition helps address the questions raised in  \cite{physrep,nlpg}.
We have applied our algorithm to compute limit coordinates in the domain in which the DHT shown
in Fig. \ref{fig:dhtqgnab}a) exists. In particular we have applied it with the input:
$\mathcal{D}=(55, 75)\times (1325, 1375)$ km$^2$,  $t_0=120$, $t_N=300$, $\Delta t=5$ days, $\tau_0=2$ days, $\Delta \tau=5$ days, and 
$\delta=10^{-3}$ km. The time step of the Runge-Kutta method is $0.1$ days. 
Fig. \ref{fig:dhti}a) indicates with a solid line
the projection onto the $x-y$ plane of the trajectory depicted in figure \ref{fig:dhtqgnab}a)
in the interval $(120, 300)$, and with circles the path of limit coordinates. Fig. \ref{fig:dhti}b) 
shows the evolution of the distances between these trajectories. This confirms that the trajectory displayed in Fig.
\ref{fig:dhtqgnab}a) is also distinguished in the sense of definition \ref{def:dtdef}
in the time interval $[120, 330]$ with accuracy $\epsilon=8 \cdot 10^{-1}$ km. 
Thus in this time interval, limit coordinates give
a method for computing DT  different from those proposed in \cite{nlpg,kayo}. Circles in Fig. \ref{fig:dhtiellhyp}
show the location  versus time 
of the $x$ limit coordinates computed with our algorithm. The solid line represents
a trajectory obtained after 
integrating with  a 5th order Runge-Kutta method  forwards and backwards in time the initial condition of the circle
at day 285. The dashed line represents the same, but with the initial condition slightly perturbed.
It is evident that in both cases the trajectories are aligned with the path of limit coordinates.
The distinguished trajectory is highly hyperbolic backwards in time as in that
direction  a small perturbation
 amplifies greatly, while it does not do so forwards in time, suggesting that  it has a non-hyperbolic type of stability
in that direction (see comments to Figs. \ref{fig:duffingtray} and \ref{fig:duffingetray}).

Beyond day 300 it is possible to continue the path of limit coordinates. Figure 
\ref{fig:dhtifin1} shows a diagram at day 330;  there  is shown the convergence of the
$x$ component of the minimum of $M$ versus $\tau$. This type of convergent diagram is not found 
in this neighbourhood for day 337. On the other hand,  although it is possible to continue the path of limit coordinates
beyond day 300, Fig. \ref{fig:dhtifin2} proves that this path is not a trajectory. There can be seen
the existence of different trajectories crossing the path, confirming that it is not a trajectory
as otherwise it would violate the uniqueness of the solution. Therefore, following our construction
it is possible to say that beyond day 300 the trajectory is no longer distinguished.

Fig. \ref{fig:dhtd} confirms that the trajectory 
in Fig. \ref{fig:dhtqgnab}b) is also distinguished in the sense of definition \ref{def:dtdef}
in the time interval $(470, 860)$ with accuracy $\epsilon=3$ km.
In particular to compute the path in Fig. \ref{fig:dhtd} we have applied the  algorithm of section
3 with the input:
$\mathcal{D}=(50, 65)\times (1255, 1270)$ km$^2$,  $t_0=470$, $t_N=860$, $\Delta t=5$ days, $\tau_0=2$ days, $\Delta \tau=7$ days, and 
$\delta=10^{-3}$ km. The Runge-Kutta time step is $0.1$ days. In the time interval from day 600 to day 650 
some of the input parameters were
 modified as follows: $\mathcal{D}=(73.5, 75.5)\times (1384, 1392)$ km$^2$, $\tau_0=40$ days, and $\Delta t=1$ day.
This  was due to the presence of nearby elliptic type minima in the function $M$, that made
difficult  tracking  the path of the limit coordinates with the previous input.   

Finally, we discuss the existence of non-hyperbolic distinguished trajectories in this data set.
The presence of this type of trajectories has not been addressed before, and  we do not have any benchmark solution. 
We have looked for this type of trajectory
in areas of the flow where Eulerian eddies seemed to persist for long times. Figure \ref{fig:Me}
represents the function $M$ at day 370  for
$\tau=150$ and $\tau=250$. In these figures there can be seen the structure of an eddie
 at the centre even for rather long $\tau$-values, however    figure \ref{fig:xytau370e} does not confirm the convergence
of the minimum of $M$ towards a constant value. On the other hand, the slow convergence
in diagram \ref{fig:duffingexy} towards the non-hyperbolic trajectory, already suggested that long time intervals
were required for that purpose, and those intervals might be difficult to find in realistic flows such 
like the one analysed here, in which one is provided just with a finite time datat set.

\section{Conclusions}

In this paper we have proposed a new definition of distinguished trajectory that
attempts to extend the concept of fixed point and periodic orbit to aperiodic dynamical systems.
The concept of fixed point is trivially contained in the definition. Regarding other
 especially useful trajectories  in dynamical systems, for instance periodic orbits, we have not proved 
that they fall within 
the definition in a general way, but we have  numerically verified it for  selected 2D and 3D  examples.
The definition can be  implemented numerically
and  the procedure consists in determining a path of {\it limit coordinates}.
We have analysed exact examples for the Duffing equation with known distinguished trajectories, 
both periodic an aperiodic, and we have found that the path of 
  limit coordinates coincides, to within  numerical accuracy, with  the distinguished trajectories
 and therefore those trajectories are identified also as {\it distinguished } in the framework 
of our definition. Our definition is novel with respect to previous works dealing with distinguished trajectories,
because it is applicable to non-hyperbolic trajectories. In particular we have studied 
a periodic orbit of the Duffing equation with non-hyperbolic  stability and is also recognised as 
{\it distinguished} by our definition. In this case the function $M$ from which 
the {\it limit coordinates} are computed seems to be a suggestive tool for characterising Lagrangian eddies.
We have tested our definition in the context of realistic aperiodic flows where 
distinguished hyperbolic trajectories had been found \cite{physrep,nlpg}. Again we have identified these trajectories
by paths of limit coordinates in certain time intervals. Beyond these time intervals the trajectories
are no longer {\it distinguished} according to our definition. Thus in the context of the
definitions provided in this paper, the property of a trajectory of being {\it distinguished} may be lost in time.
Also we have found evidence  that the hyperbolicity of
these trajectories is not  constant in time. These two statements
provide answers to the open questions mentioned in the text that have been addressed in \cite{physrep,nlpg}.

\section*{Acknowledgements}
The authors have been supported by CSIC grants PI-200650I224 and OCEANTECH (No. PIF06-059), Consolider I-MATH (C3-0104)
and the Comunidad de Madrid project SIMUMAT S-0505-ESP-0158. 
 It is a pleasure to acknowledge many
useful conversations with Steve Wiggins, Des Small, Peter Haynes,  Emilio Hern\'andez-Garc\'{\i}a, Crist\'obal L\'opez, Antonio Turiel,
Emilio Garc\'{\i}a-Ladona, Michal Branicki, Wenbo Tang,  J.J. L. Vel\'azquez on numerous issues  related to this project.
We also thank very valuable suggestions from Daniel Fox, Andrew Thompson, 
and Jodie Holdway. 
The computational part of this work was done using the CESGA computers SVGD and FINIS TERRAE and using the
SIMUMAT-CSIC cluster ODISEA.

\newpage

{\bf Figure captions}

Figure 1

Solutions (\ref{sol1}) for different initial conditions $x^*$. a) Solutions for positive times $t>0$. b) Solutions for   positive and negative times.

Figure 2

a) Function $M(x^*)_{t=0,\tau}$ evaluated over the solutions (\ref{sol1}). Dashed line $\tau=3$, solid line $\tau=4$. b) Position of 
the $x^*$-coordinate at the minimum of the function $M_{t=0,\tau}$ as a function of $\tau$. The horizontal dashed line marks the position of the DHT.

Figure 3

Position of 
the $x^*$-coordinate at the minimum of the function $M_{t=0,\tau}$  as a function of $\tau$.  The function $M_{t=0,\tau}$ is considered for the solutions in 
(\ref{sol2}). The horizontal dashed line marks the position of the DHT.

Figure 4

Contour plot of the function $M_{t=0, \tau=2}({\bf x})$ in the open set
${\mathcal D} =(-0.2,0.2)\times(-0.2,0.2)$. The minimum corresponds to the black tone.

Figure 5

Contour plot of the function $M$ in the open set
${\bf x}\in (-0.2,0.2)\times(-0.2,0.2)$. a) $M_{t=0, \tau=5}({\bf x})$; b) $M_{t=0, \tau=10}({\bf x})$; c)
$M_{t=0, \tau=50}({\bf x})$.

\begin{figure}[htb]
\vspace{-6cm}
\begin{center}
\includegraphics[width=1.2\columnwidth]{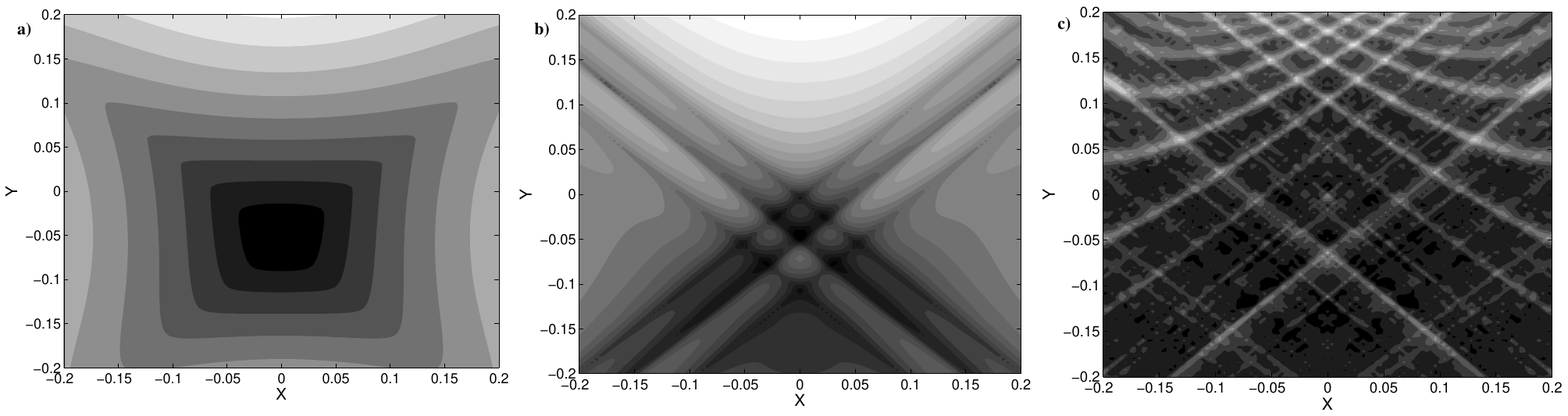}
\end{center}
\caption{}\label{fig:duffingtau}
\end{figure}

\begin{figure}[htb]
\vspace{-6cm}
\begin{center}
\includegraphics[width=1.2\columnwidth]{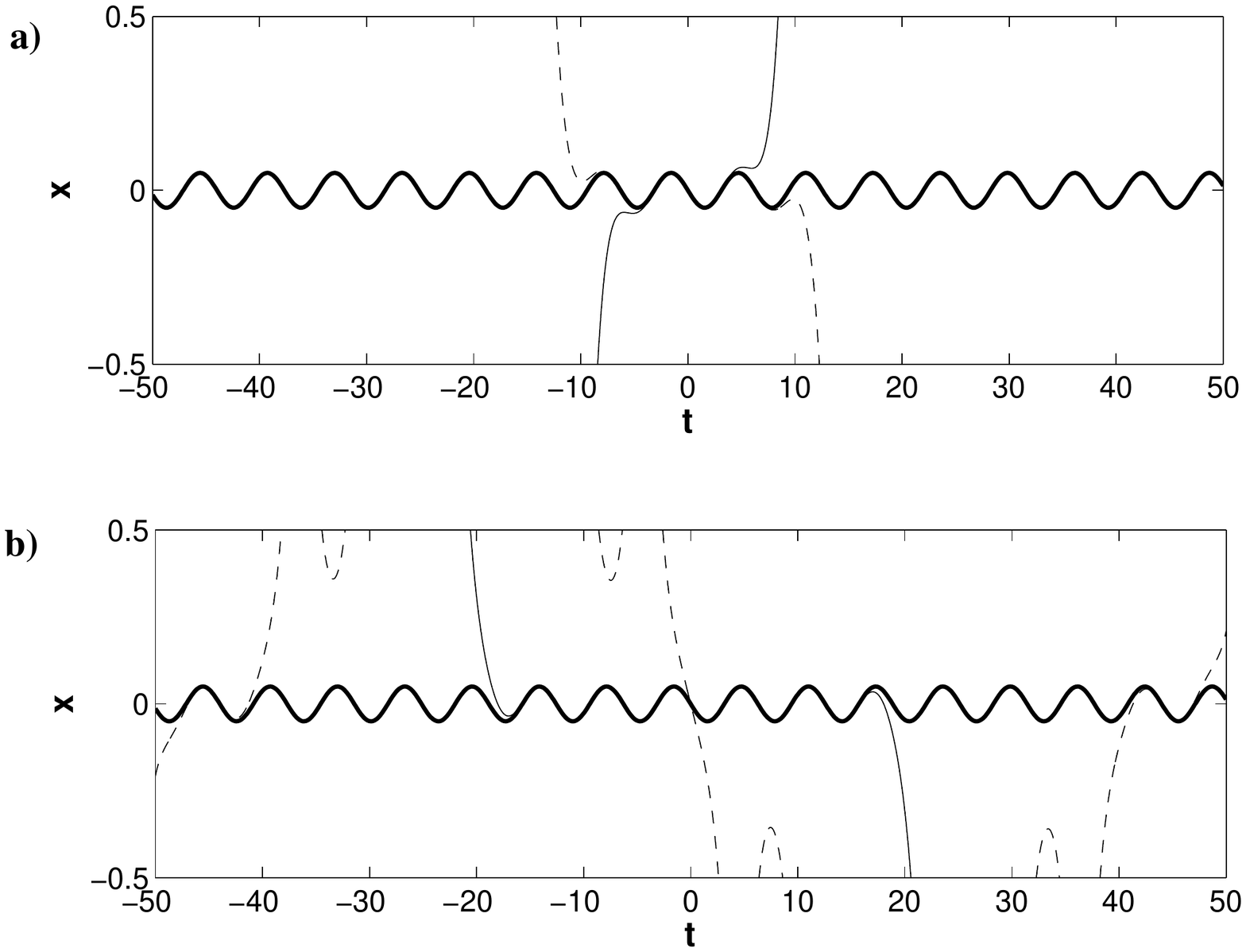}
\end{center}
\caption{}\label{fig:duffingtray}
\end{figure}

\begin{figure}[htb]
\vspace{-6cm}
\begin{center}
\includegraphics[width=1.2\columnwidth]{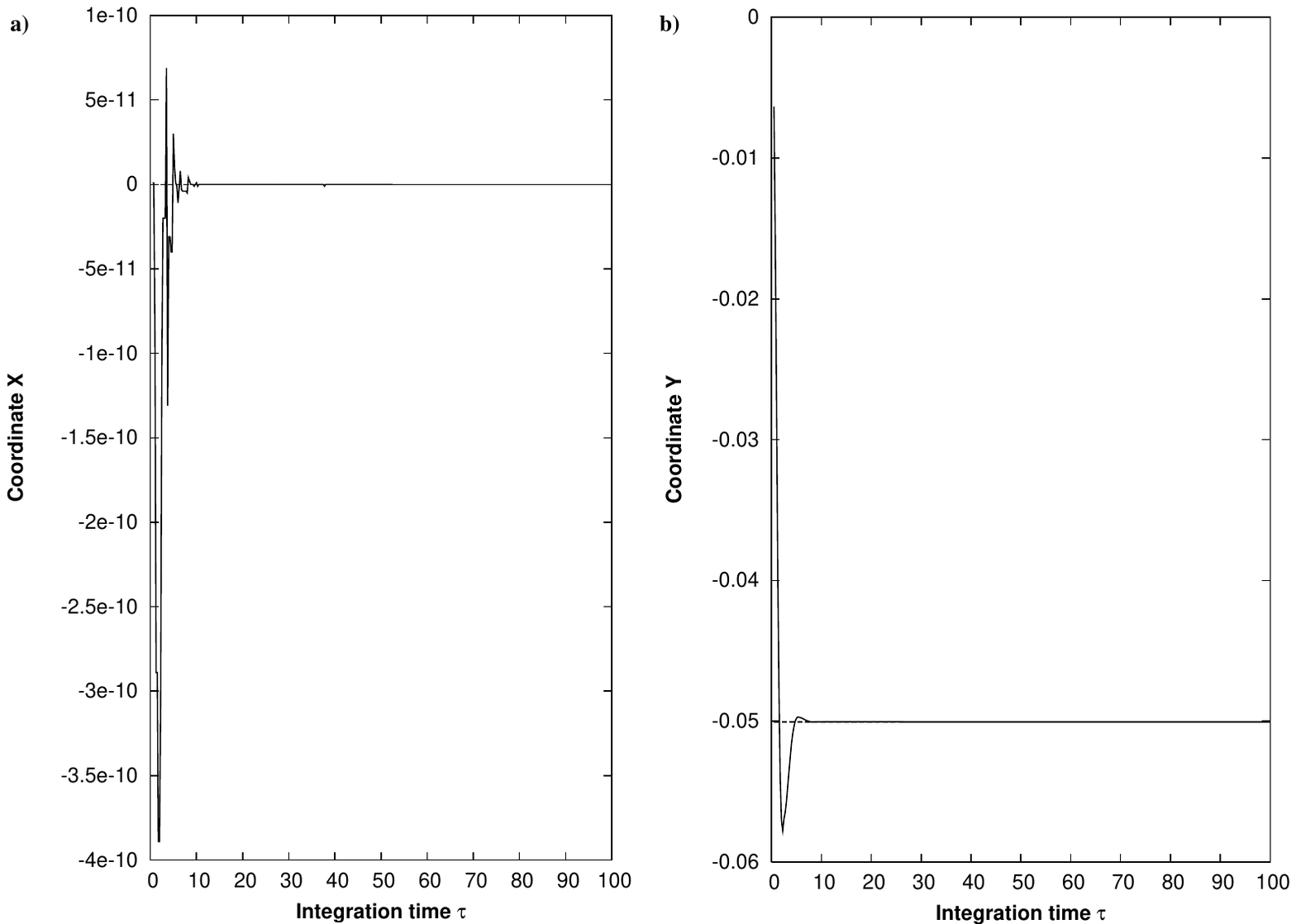}
\end{center}
\caption{}\label{fig:duffingxy}
\end{figure}

\begin{figure}[htb]
\begin{center}
\includegraphics[width=1.1\columnwidth]{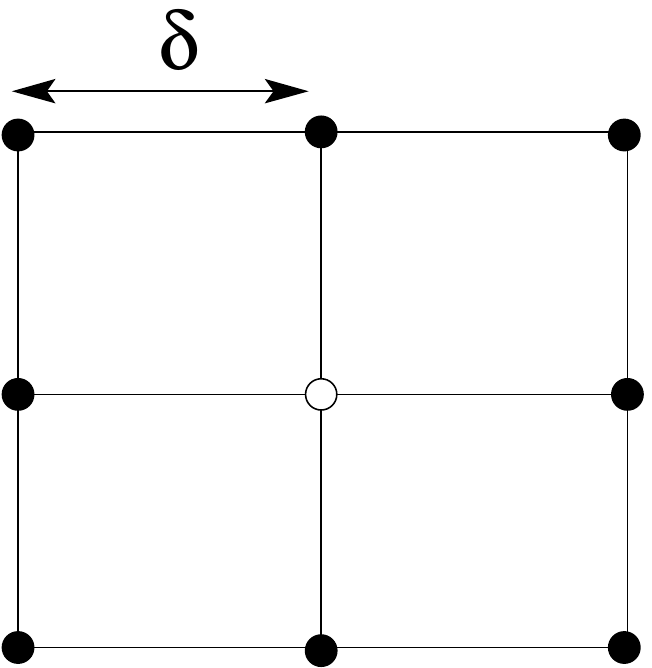}
\end{center}
\caption{}\label{fig:epsilongrid}
\end{figure}

\begin{figure}[htb]
\begin{center}
\includegraphics[width=1.1\columnwidth]{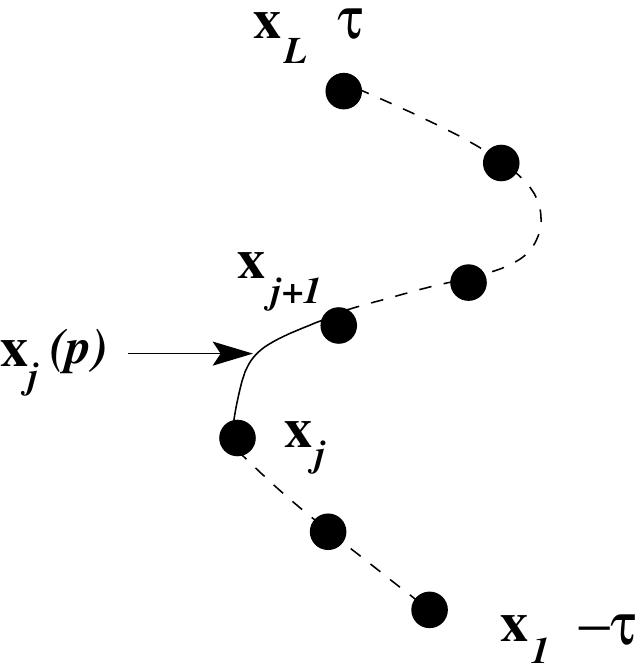}
\end{center}
\caption{}\label{tra1}
\end{figure}

\begin{figure}[htb]
\begin{center}
\includegraphics[width=1\columnwidth]{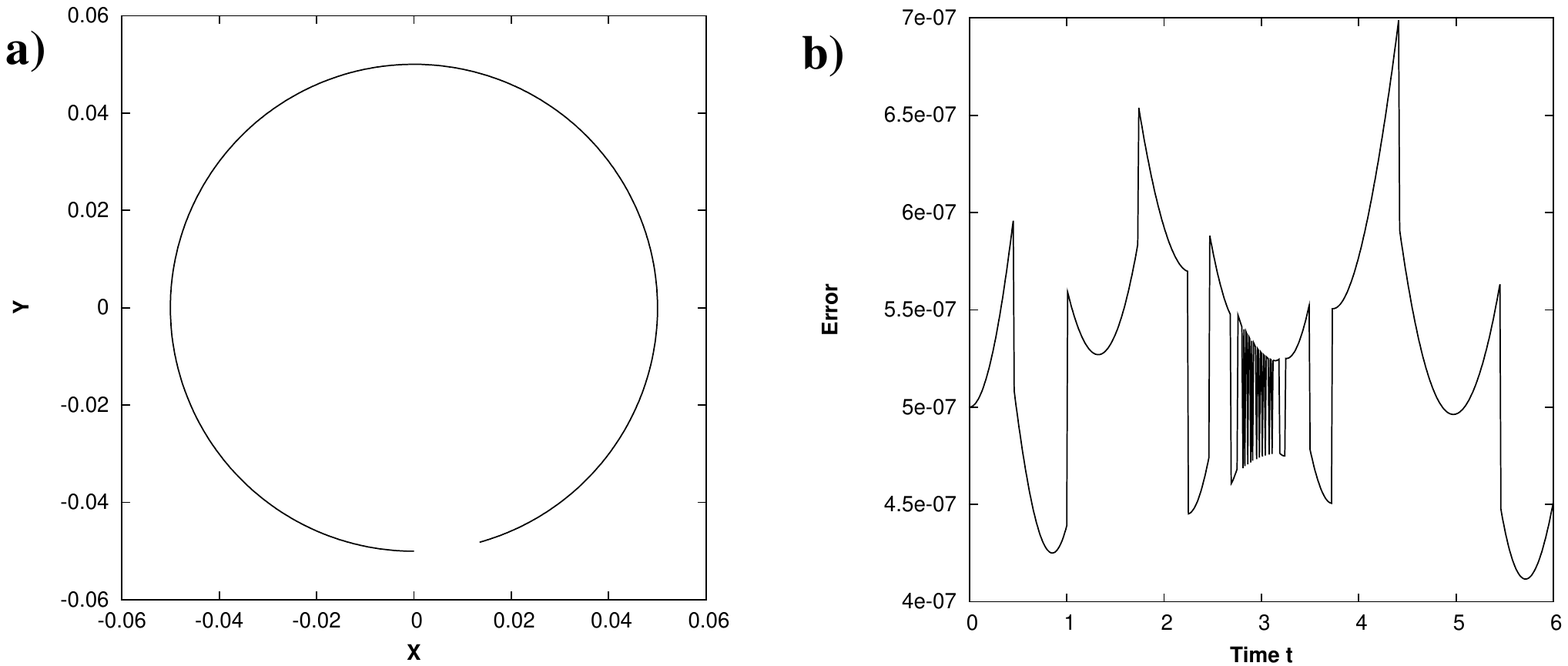}
\end{center}
\caption{}\label{fig:dhtduffing}
\end{figure}

\begin{figure}[htb]
\begin{center}
\includegraphics[width=0.9\columnwidth]{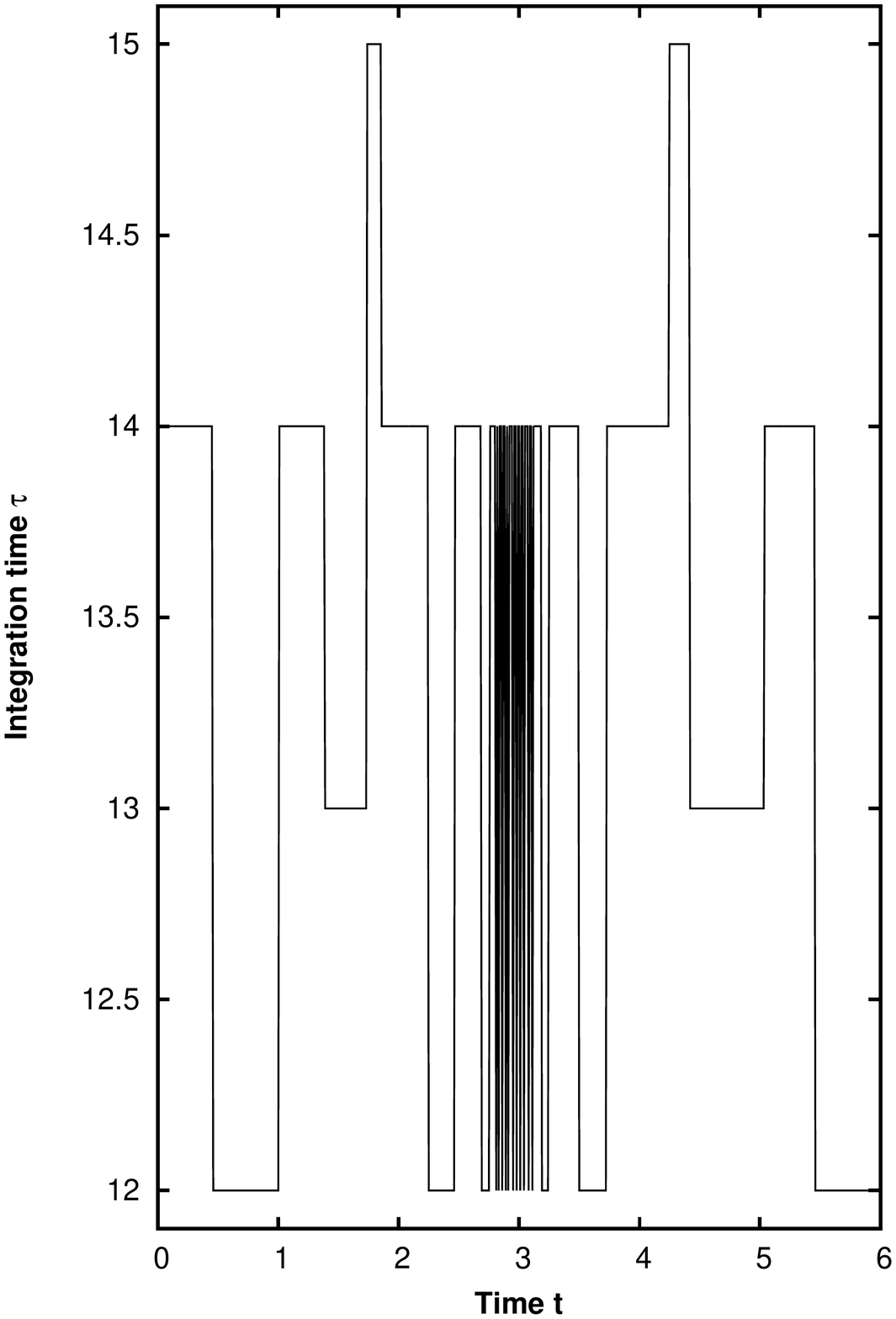}
\end{center}
\caption{}\label{fig:dufftau}
\end{figure}

\begin{figure}[htb]
\begin{center}
\includegraphics[width=1.2\columnwidth]{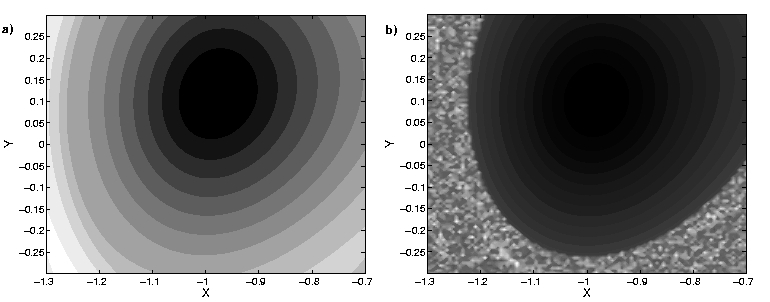}
\end{center}
\caption{}\label{fig:duffingetau}
\end{figure}

\begin{figure}[htb]
\vspace{-5cm}
\begin{center}
\includegraphics[width=1.1\columnwidth]{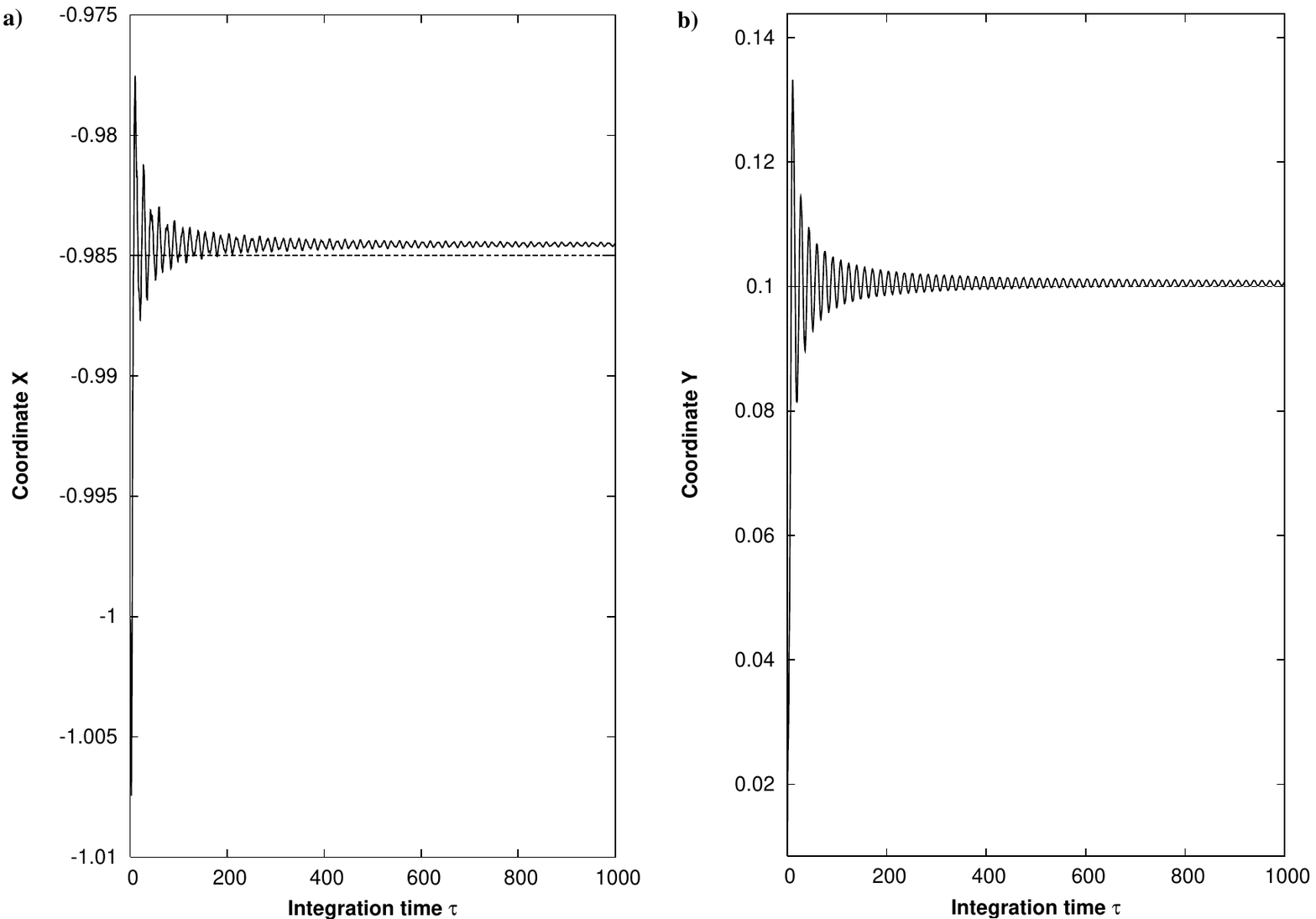}
\end{center}
\caption{}\label{fig:duffingexy}
\end{figure}

\begin{figure}[htb]
\vspace{-4cm}
\begin{center}
\includegraphics[width=1.2\columnwidth]{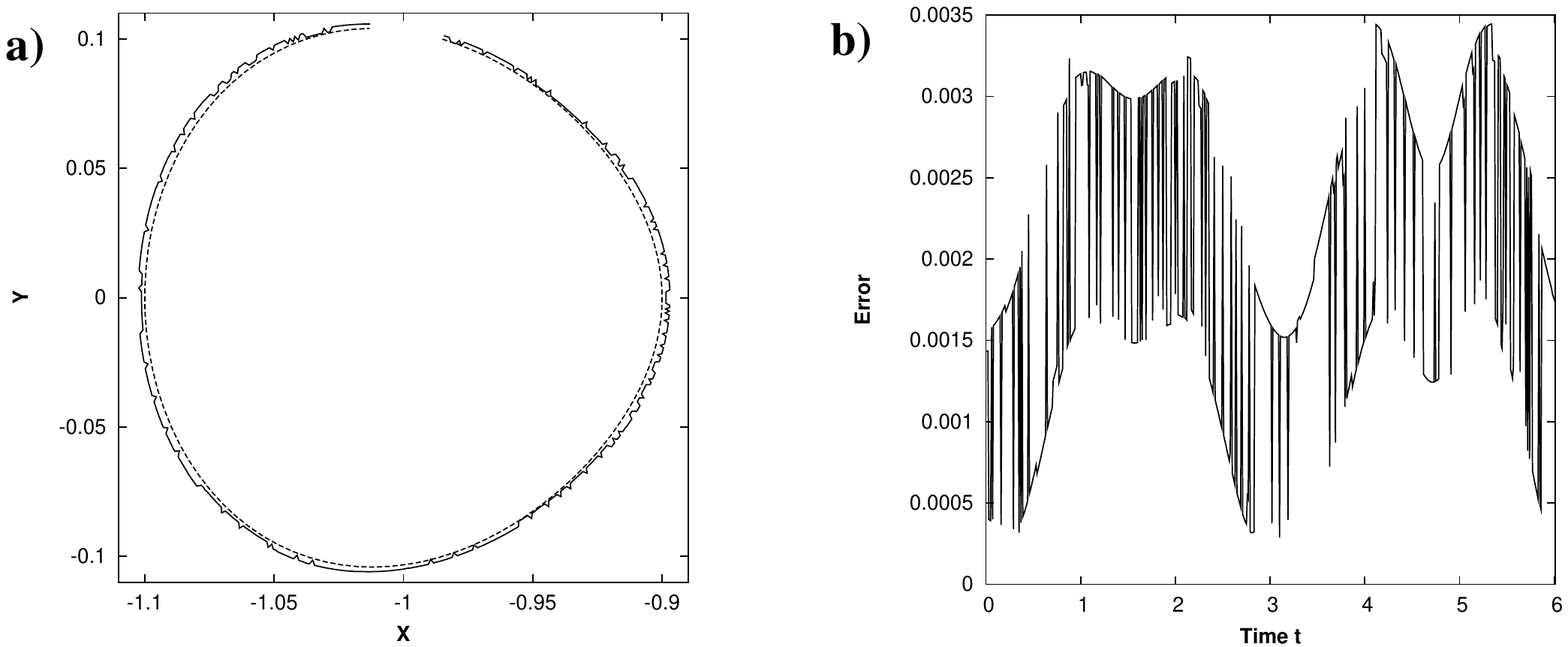}
\end{center}
\caption{}\label{fig:detduffing}
\end{figure}

\begin{figure}[htb]
\vspace{-4cm}
\begin{center}
\includegraphics[width=1.3\columnwidth]{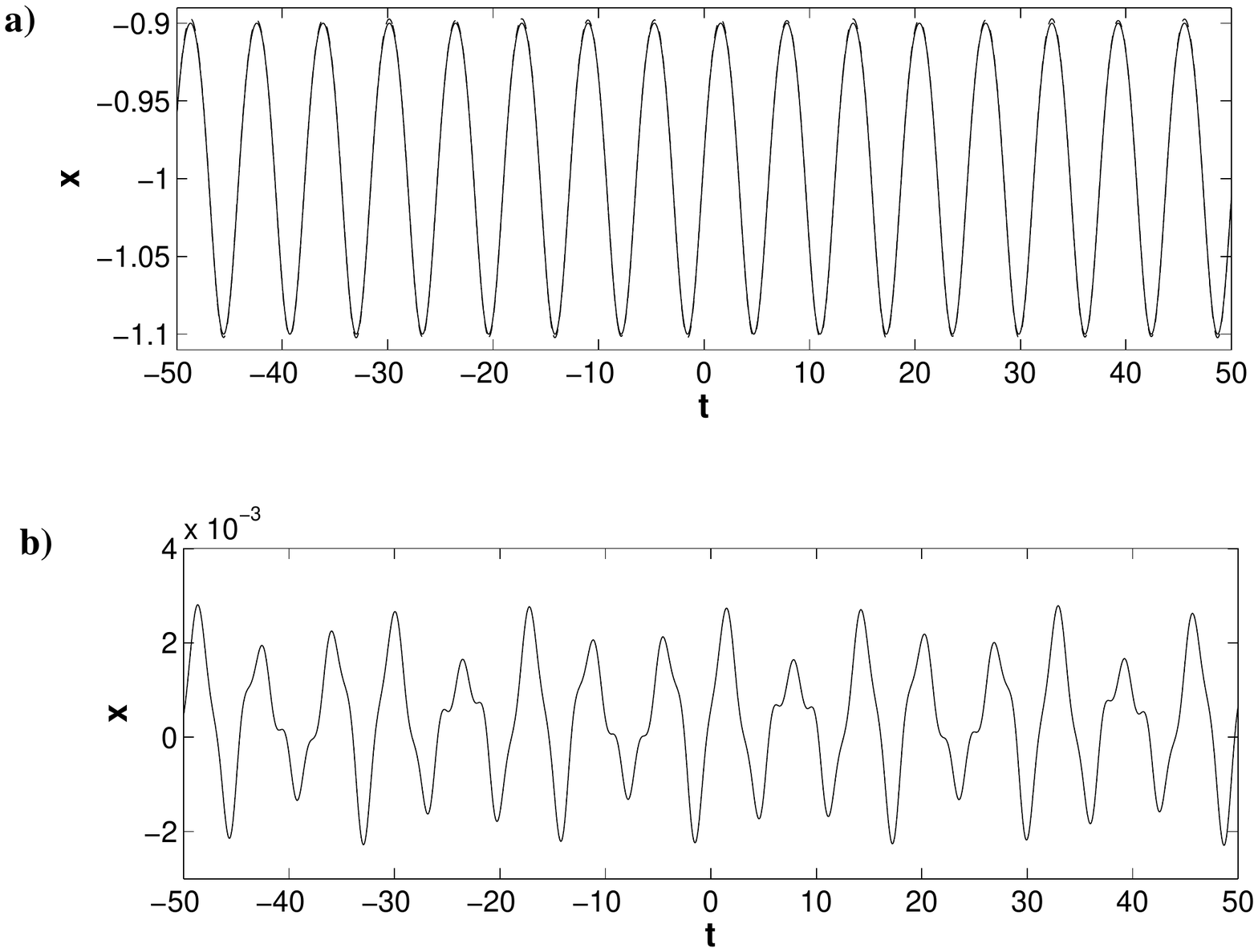}
\end{center}
\caption{}\label{fig:duffingetray}
\end{figure}

\begin{figure}[htb]
\begin{center}
\includegraphics[width=1.2\columnwidth]{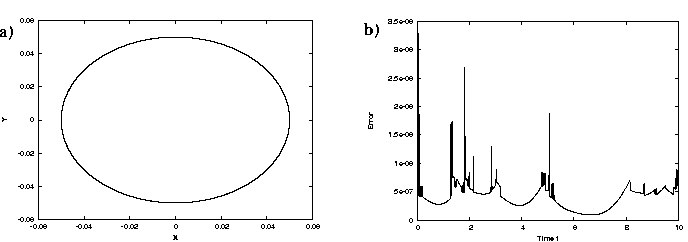}
\end{center}
\caption{}\label{fig:dhtrduffing}
\end{figure}

\begin{figure}[htb]
\vspace{-4cm}
\begin{center}
\includegraphics[width=0.8\columnwidth]{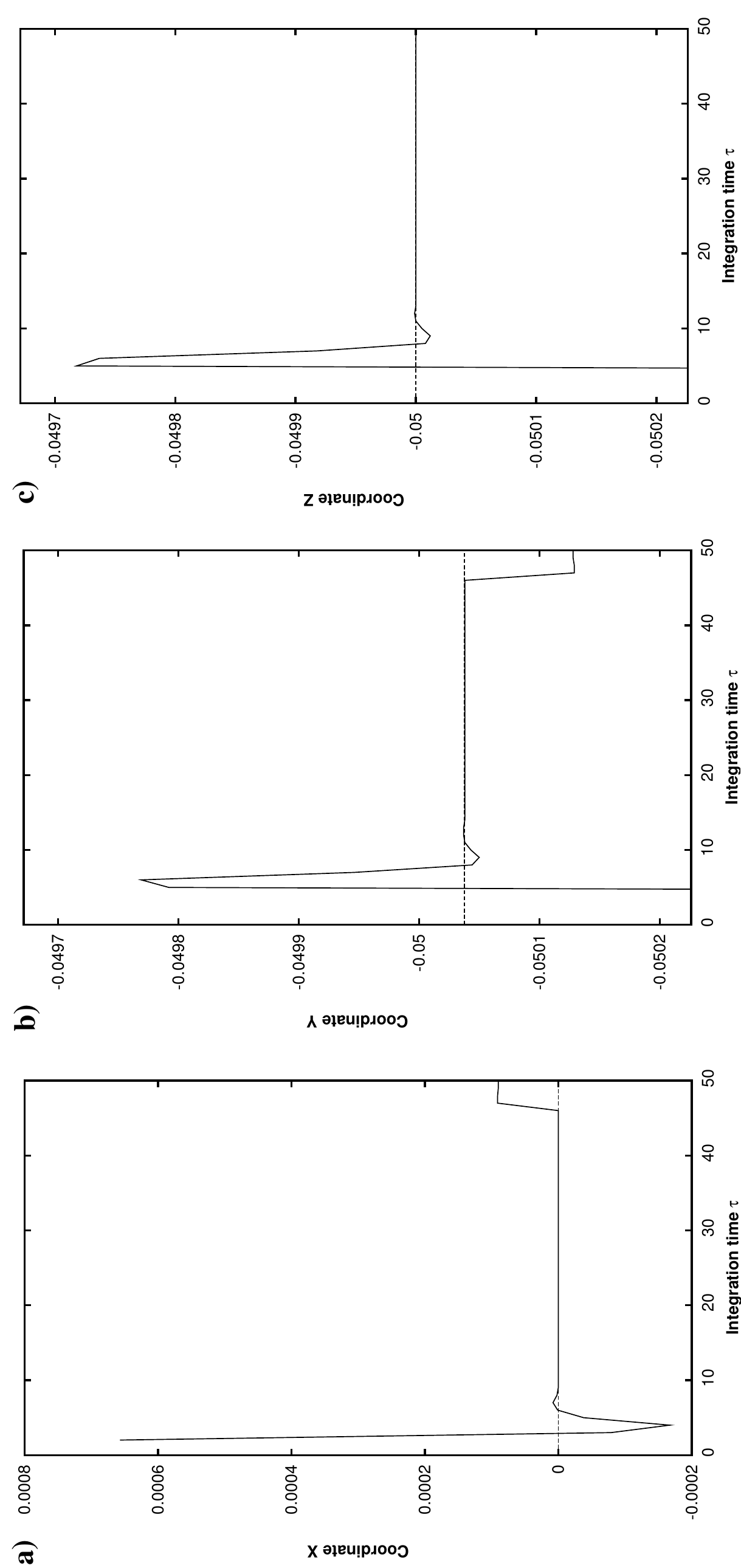}
\end{center}
\caption{}\label{fig:duffingtau3d}
\end{figure}

\begin{figure}[htb]
\vspace{-14cm}
\begin{center}
\includegraphics[width=1.8\columnwidth]{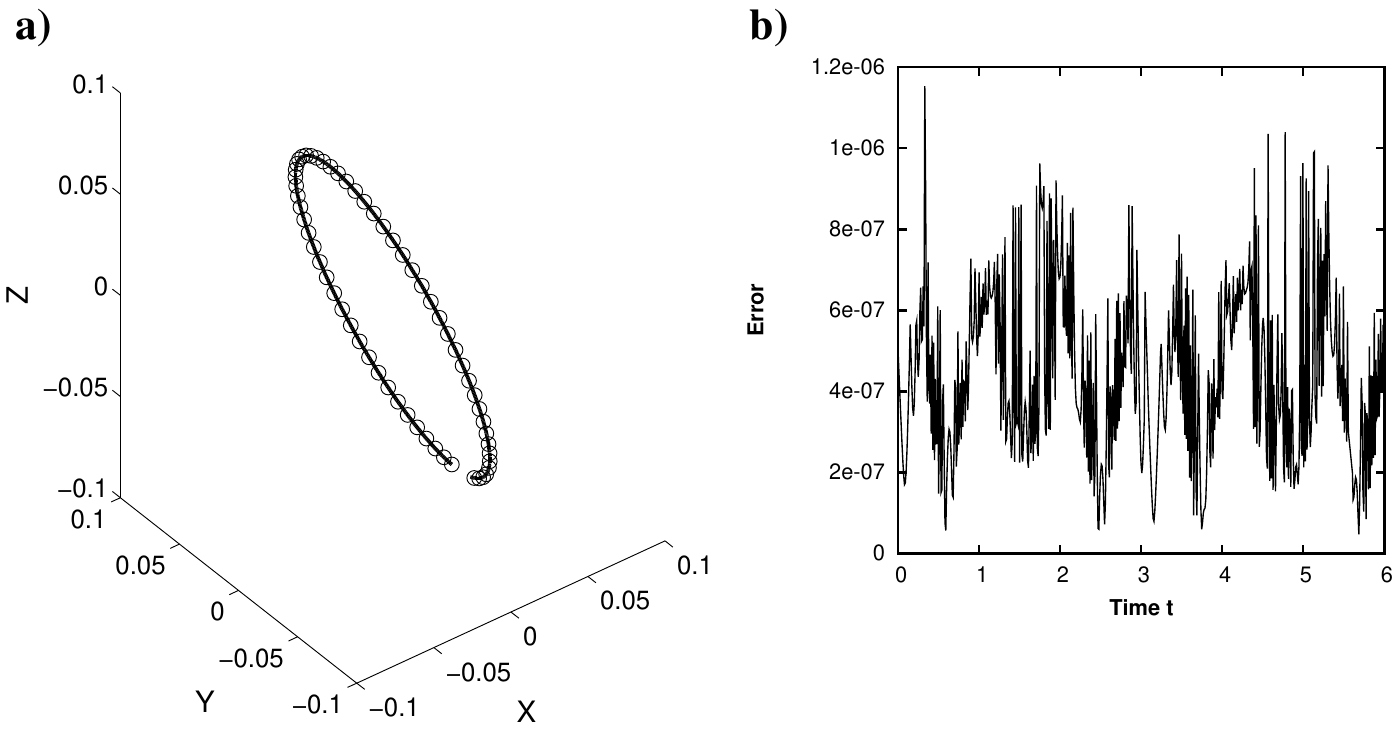}
\end{center}
\caption{}\label{fig:duffing3d}
\end{figure}

\begin{figure}[htb]
\includegraphics[width=1.4\columnwidth]{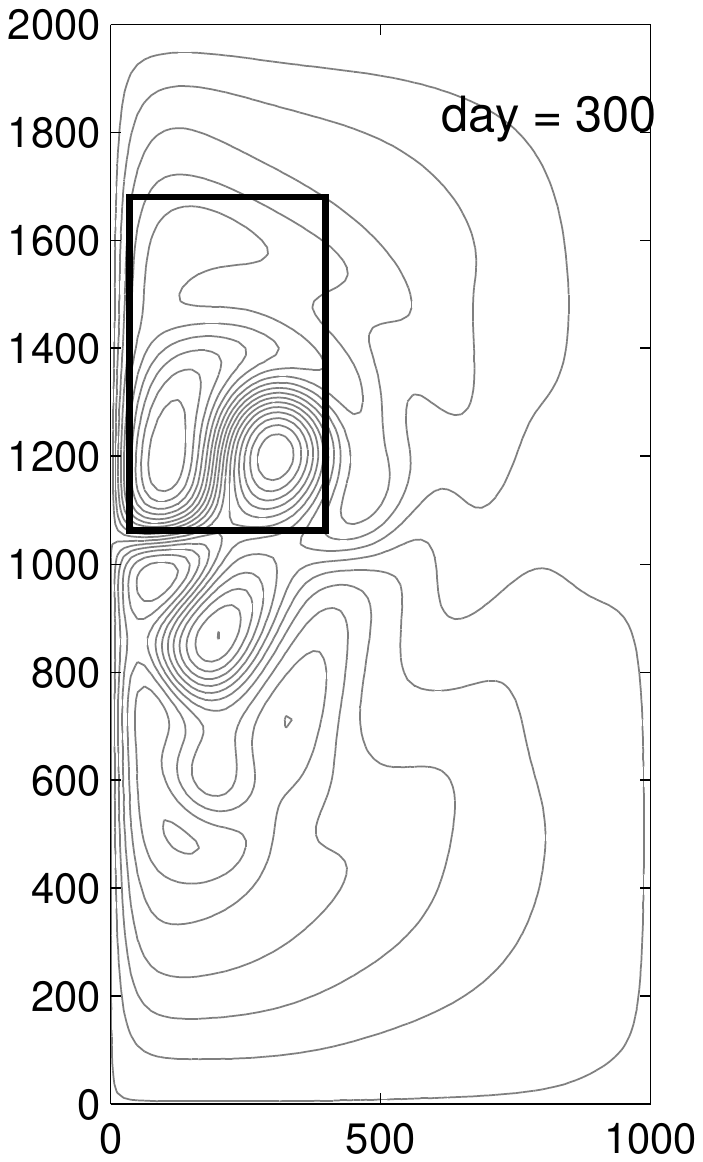}
\vspace{-6cm}
\caption{}\label{fig:sfqgn}
\end{figure}

Figure 6

a) $x$-coordinate versus time for the DHT (thick solid line) and those trajectories integrated
with initial conditions  at the global minima of Figs.  \ref{fig:duffingtau}a) (solid line) and b) (dashed line); b) $x$-coordinate versus time for the DHT (thick solid line),  a trajectory integrated
with initial condition  at the global minimum of Fig.  \ref{fig:duffingtau}c) (solid line) and a trajectory integrated
at a non-global but relative minimum of the same figure (dashed line).

Figure 7

Evolution of the  coordinates of the global minimum of $M$  
versus  $\tau$. a) The $x$ coordinate; b) the $y$ coordinate.
 These plots show the convergence to the
DHT whose position is marked with a dashed horizontal line.

Figure 8

A $\delta$-grid in $\mathbb{R}^2$. The centre or interior point
 is marked with the white dot.

Figure 9

A schematic projection onto the $\mathbb{R}^2$ plane of a possible trajectory from $-\tau$ to $\tau$ with $L$ points.

Figure 10

a) Representation  of both the 
distinguished hyperbolic trajectory  (\ref{eq:dhtduffing})  and its
approximation obtained with the proposed numerical algorithm for $\varepsilon=0.1$; b)  distance between the 
exact and the numerical approach.

Figure 11

Representation of the maximum $\tau$ required to approach the DT to within accuracy $\delta=10^{-6}$ versus time.

Figure 12

Contour plot of the function $M$ in the open set
${\bf x}\in (-1.2,-0.8)\times(-0.2,0.2)]$. a) $M_{t=0, \tau=10}({\bf x})$; b) $M_{t=0, \tau=300}({\bf x})$.

Figure 13

Evolution of the  coordinates of the global minimum of $M$  
versus  $\tau$ at $t_0=0$. a) The $x$ coordinate; b) the $y$ coordinate.
 These plots show the convergence of the minima to the
coordinates of the DET whose position is marked with a continuous horizontal line.

Figure 14

a) Dotted line represents the exact non-hyperbolic  distinguished trajectory
and the solid line stands for the numerically computed limit coordinates; b) distance between the 
exact non-hyperbolic trajectory (\ref{eq:detduffing}) and the limit coordinates.

Figure 15

a) $x$-coordinate versus time for the DET (solid line) and the trajectory integrated 
taking as initial data  the limit coordinates located at time $t_0=0$ (dashed line); b)
time evolution of the differences between these trajectories.

Figure 16

a) Dashed line represents the exact   distinguished hyperbolic trajectory of the rotating Duffing equation
and the solid line stands for the numerically computed one; b) distance between the exact and the numerical distinguished hyperbolic trajectories.

Figure 17

Evolution of the  coordinates of the global minimum of $M$  for the 3D example
versus  $\tau$ at $t_0=0$. a) The $x$ coordinate; b) the $y$ coordinate; c) the $z$ coordinate.
 These plots show the convergence of the minima to the
coordinates of the DHT whose position is marked with a dashed horizontal line.

Figure 18

a) The solid line represents the exact   distinguished hyperbolic trajectory of the 3D equation
and circles stand for numerically computed coordinates; b) distance between the exact and the numerical distinguished hyperbolic trajectories.

Figure 19

Contour plot of the streamfunction produced by the quasigeostrophic model at day 300.

Figure 20

Distinguished hyperbolic trajectories in the Northern gyre of the quasigeostrophic model
reported in \cite{nlpg}. a) Evolution of the $x$ and $y$ coordinates in the time interval $[5, 338]$;
b) evolution of the $x$ and $y$ coordinates in the time interval $[450, 880]$ 

\begin{figure}
\includegraphics[width=1.1\columnwidth]{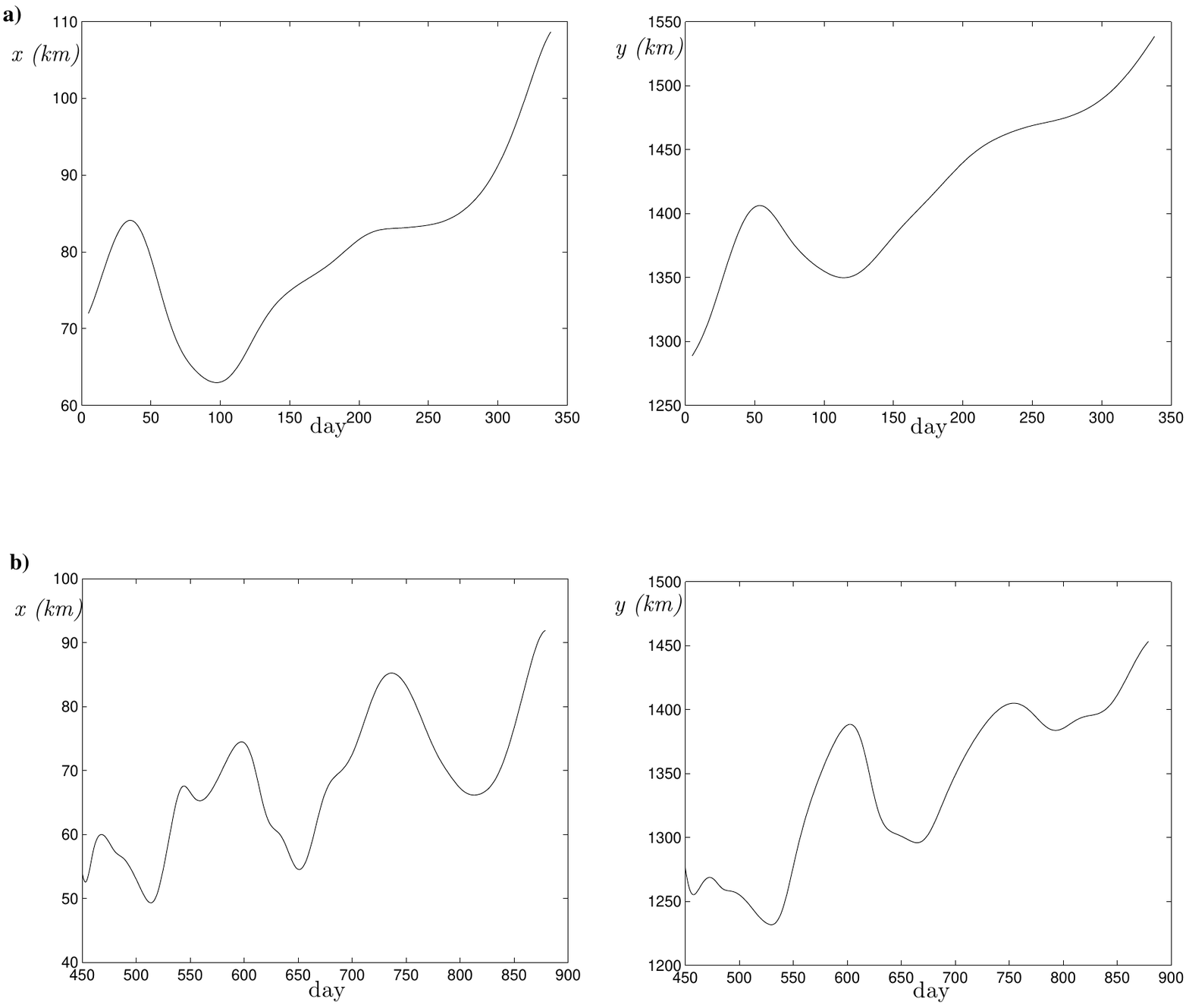}
\caption{}\label{fig:dhtqgnab}
\end{figure}

\begin{figure}
\vspace{-4cm}
\includegraphics[width=1.2\columnwidth]{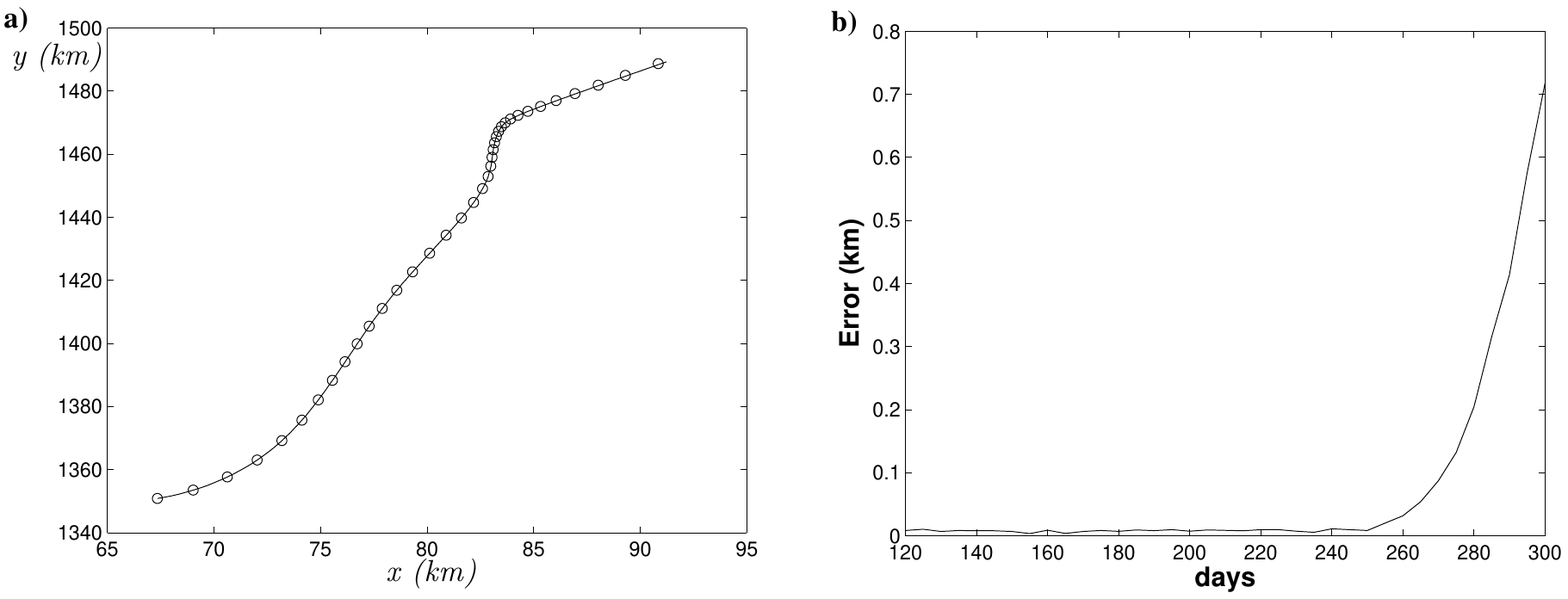}
\caption{}\label{fig:dhti}
\end{figure}

\begin{figure}
\includegraphics[width=1.2\columnwidth]{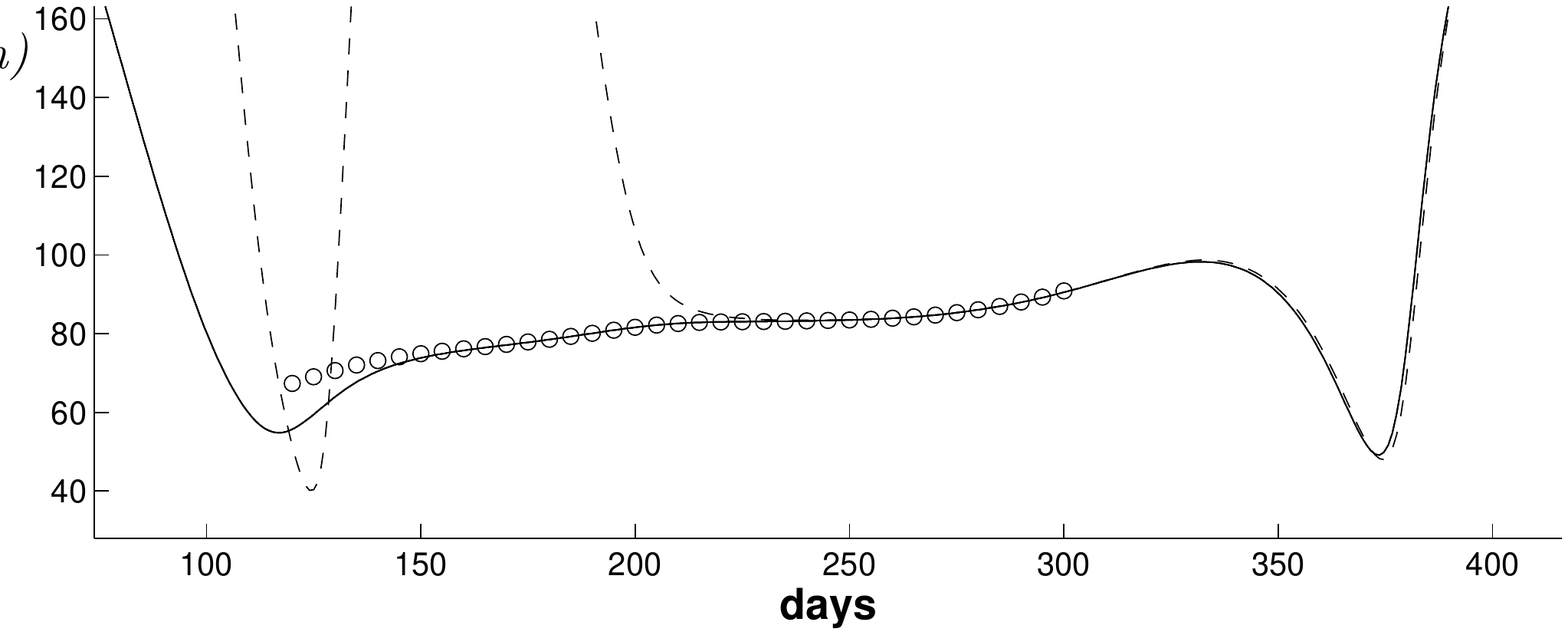}
\caption{}\label{fig:dhtiellhyp}
\end{figure}

\begin{figure}
\includegraphics[width=1.1\columnwidth]{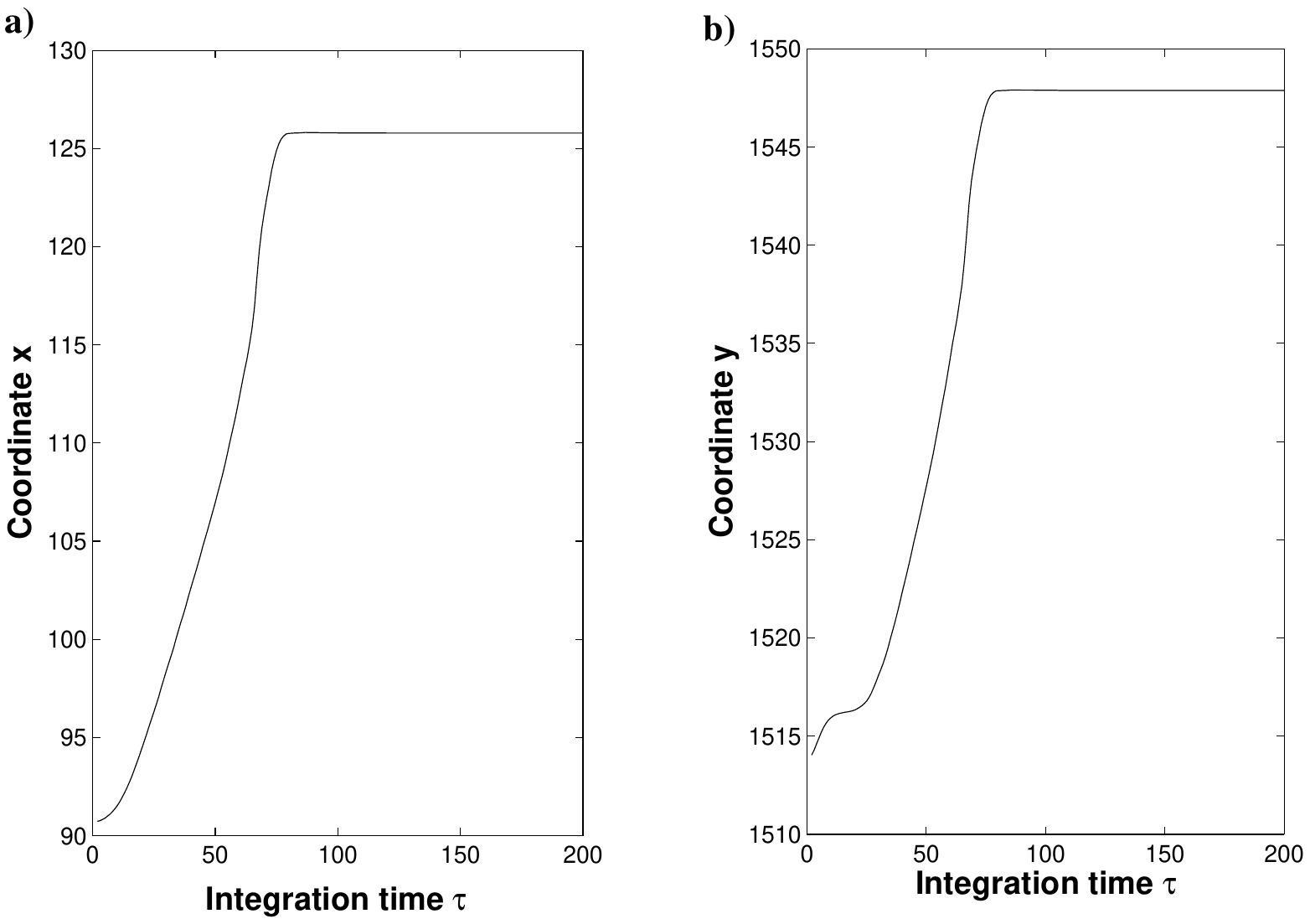}
\caption{}\label{fig:dhtifin1}
\end{figure}

\begin{figure}
\vspace{-14cm}
\hspace{-2cm}
\includegraphics[width=1.8\columnwidth]{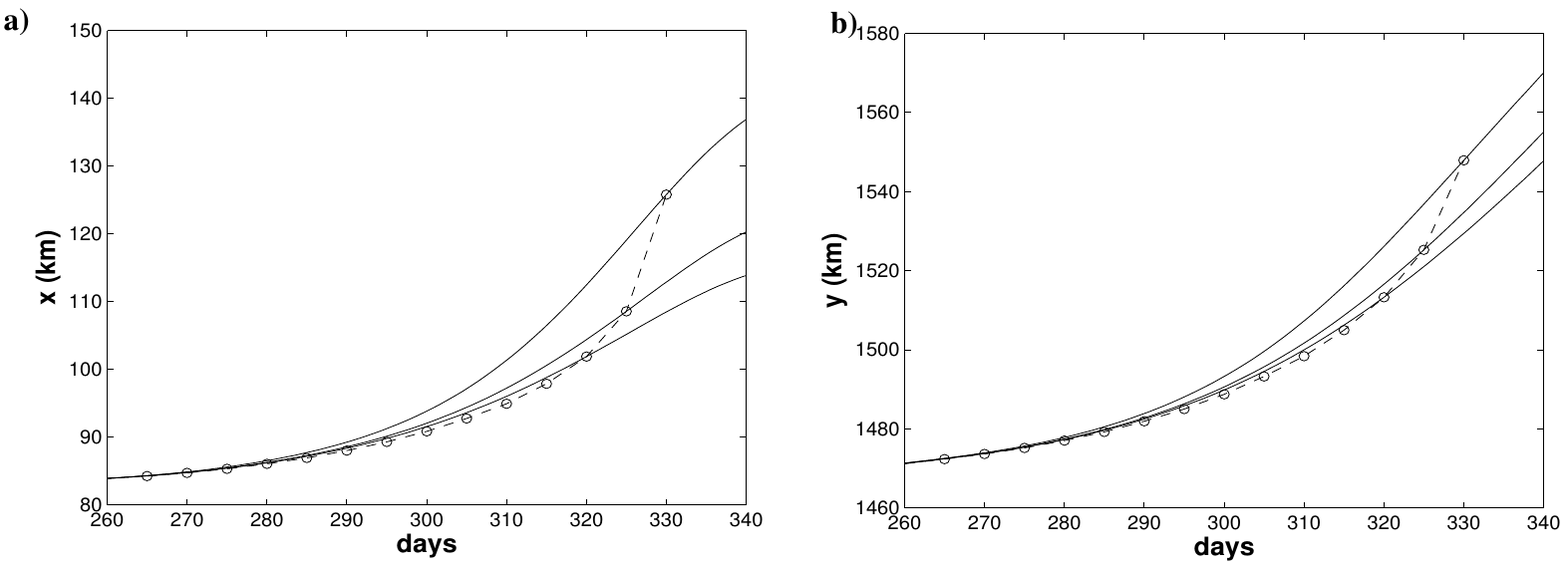}
\caption{}\label{fig:dhtifin2}
\end{figure}

\begin{figure}
\begin{center}
\vspace{-14cm}
\hspace{-2.5cm}
\includegraphics[width=1.7\columnwidth]{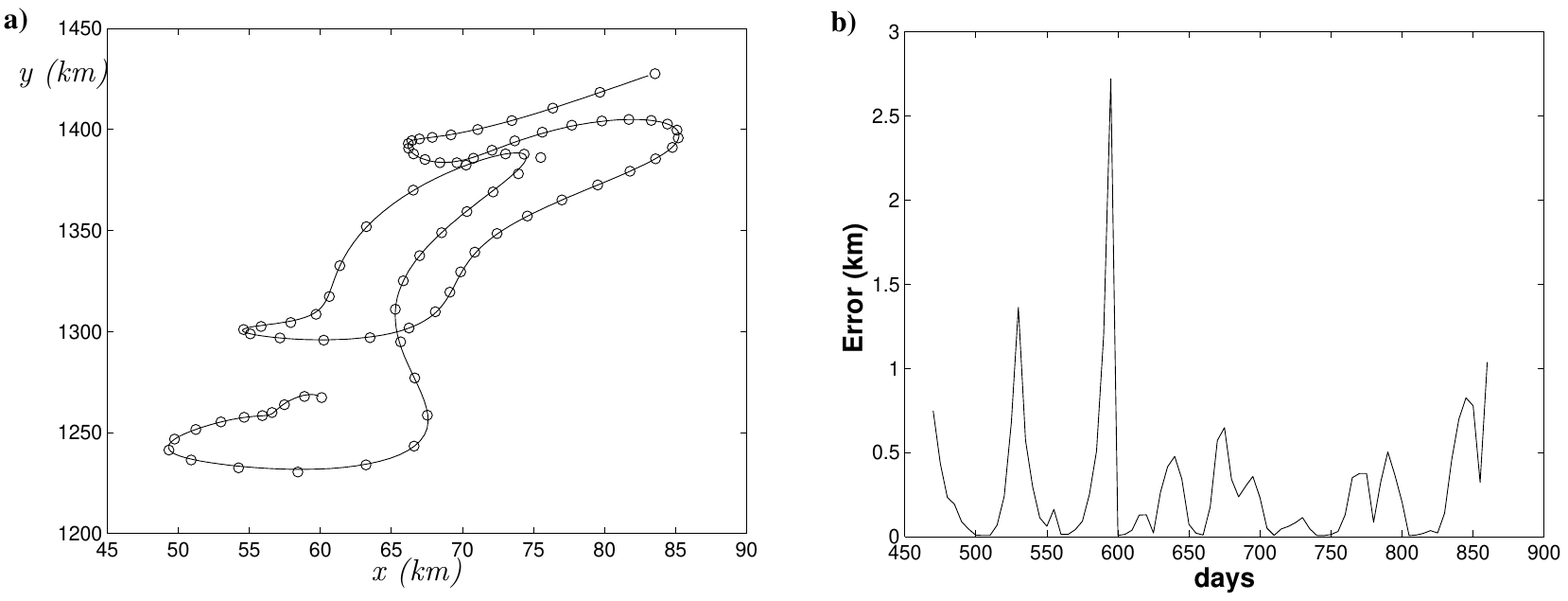}
\end{center}
\caption{}\label{fig:dhtd}
\end{figure}

\begin{figure}
\begin{center}
\includegraphics[width=1.1\columnwidth]{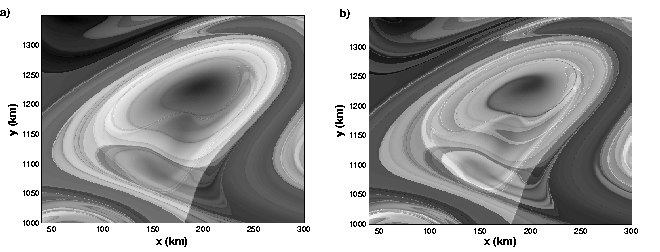}
\end{center}
\caption{}\label{fig:Me}
\end{figure}

\begin{figure}
\vspace{-14cm}
\hspace{-2.5cm}
\begin{center}
\includegraphics[width=1.7\columnwidth]{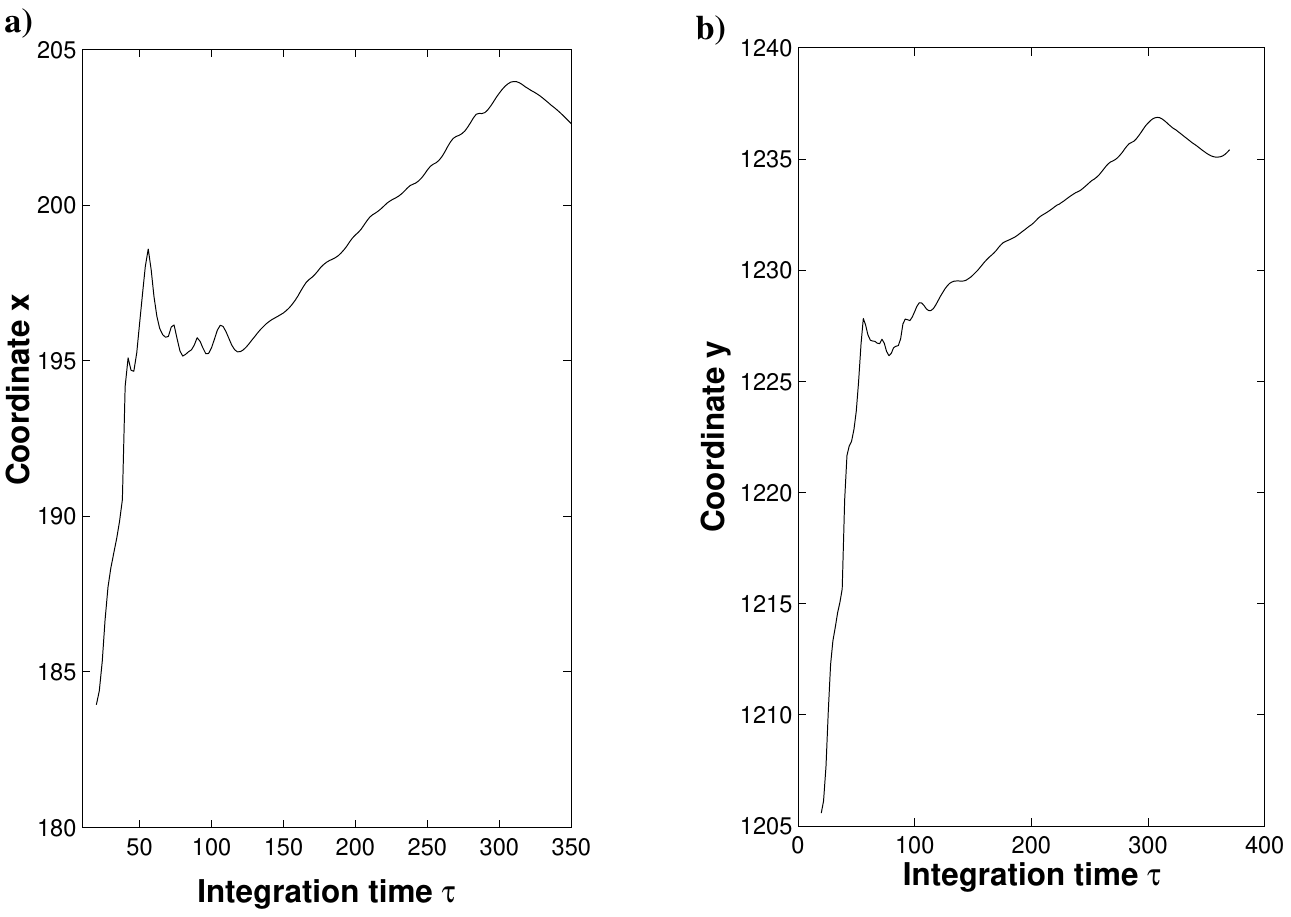}
\end{center}
\caption{}\label{fig:xytau370e}
\end{figure}

Figure 21

a) Solid line represents the  projection on the phase space of the distinguished hyperbolic trajectory depicted
in Fig. \ref{fig:dhtqgnab}a) 
and circles stands for the numerically computed limit coordinates; b) distance between 
the trajectories represented in a).

Figure 22

Circles stand for the $x$ component of the limit coordinates in the time range where they approach
a DT.  The solid line represents a trajectory integrated with a 5th order Runge-Kutta method passing the through
limit coordinates at day 285. The dashed line is a trajectory integrated from the same condition plus a small perturbation.

Figure 23

a) $x$ component of the minimum of $M$ versus $\tau$ at day 330; b) $y$ component of the minimum of $M$ versus $\tau$ at the same day.

Figure 24

a) Circles stand for the $x$ component of the limit coordinates versus time and 
the solid lines stand for different trajectories; b) the same as a) but for the $y$ component.

Figure 25

a) Solid line represents the  projection on the phase space of the  distinguished hyperbolic trajectory depicted
in Fig. \ref{fig:dhtqgnab}b) 
and circles stands for the numerically computed limit coordinates; b) distance between 
the trajectories represented in a).

Figure 26

a) Contour plot of $M_{t=370,\tau=150}$, the elliptic minimum is in the dark area almost at the centre; b) contour plot of $M_{t=370,\tau=250}$. 

Figure 27

a) $x$ component of the minimum of $M$ versus $\tau$ at day 370; b) $y$ component of the minimum of $M$ versus $\tau$ at the same day.  

{\bf Table captions}

Table 1

Relative errors for several ellipse lengths,  
computed with a linear interpolation over $L$ points on the curve.

Table 2

Relative errors for several ellipse lengths,  
computed with Dritschel interpolation over $L$ points on the curve.

\newpage

\begin{table}[htb]
\footnotesize{
\begin{flushleft}
\begin{center}
\begin{tabular}{|c||c|c|c|c|c|c|}
\hline
&\multicolumn{6}{c|}{Linear interpolation}\\
\cline{2-7}
$L$&\multicolumn{6}{c|}{Ratio between axes}\\
\cline{2-7}
&$1$&$2$&$5$&$10$&$100$&$1000$\\
\hline
\hline
$10$ & $8.16$&$8.80$ &$9.48$ &$9.73$ & $9.99$ & $10.00$\\ 
\hline    
$10^2$ &$2.63$ &$3.36$ &$3.86$&$3.99$ &$4.05$ & $4.05$\\  
\hline
$10^3$ &$0.83$  &$1.08$ &$1.24$ &$1.29$ &$1.31$ & $1.31$\\    
\hline
$10^4$ & $0.26$  &$0.34$ &$0.39$ & $0.41$ & $0.41$& $0.41$\\    
\hline
$10^5$& $8.34\times10^{-2}$&$0.11$  &$0.12$& $0.13$  &$0.13$  & $0.13$\\
\hline
$10^6$ &$2.64\times10^{-2}$&$3.42\times10^{-2}$  &$3.94\times10^{-2}$ & $4.08\times10^{-2}$ & $4.14\times10^{-2}$ &$4.14\times10^{-2}$\\
\hline
$10^7$ & $8.34\times10^{-3}$&$1.08\times10^{-2}$  &$1.25\times10^{-2}$ & $1.29\times10^{-2}$ &$1.31\times10^{-2}$ & $1.31\times10^{-2}$\\
\hline
$10^8$ & $2.64\times10^{-3}$&$3.42\times10^{-3}$  &$3.95\times10^{-3}$ &$4.08\times10^{-3}$ &$4.14\times10^{-3}$ &$4.14\times10^{-3}$\\
\hline
$10^9$ &$8.34\times10^{-4}$&$1.08\times10^{-3}$  &$1.25\times10^{-3}$ & $1.29\times10^{-3}$ & $1.31\times10^{-3}$ & $1.31\times10^{-3}$\\
\hline
\end{tabular}
\end{center}
\end{flushleft}
}
\caption{}
\label{tab:erroresl}
\end{table}

\begin{table}[htb]
\footnotesize{
\begin{flushleft}
\begin{center}
\begin{tabular}{|c||c|c|c|c|c|c|}
\hline
&\multicolumn{6}{c|}{Dritschel interpolation}\\
\cline{2-7}
$L$&\multicolumn{6}{c|}{Ratio between axes}\\
\cline{2-7}
&$1$&$2$&$5$&$10$&$100$&$1000$\\
\hline
\hline
$10$ & $0.99$&$0.67$ &$0.33$ &$0.25$ &$0.26$ & $0.27$ \\ 
\hline    
$10^2$ & $3.49\times10^{-2}$ &$1.37\times10^{-2}$ &$4.47\times10^{-3}$ &$2.81\times10^{-3}$ &$2.08\times10^{-3}$  &$2.61\times10^{-3}$\\  
\hline
$10^3$ & $1.11\times10^{-3}$ &$3.83\times10^{-4}$ &$9.20\times10^{-5}$ &$4.43\times10^{-5}$ &$2.23\times10^{-5}$ & $1.98\times10^{-5}$ \\    
\hline
$10^4$ & $3.53\times10^{-5}$ &$1.17\times10^{-5}$ &$4.44\times10^{-6}$ &$5.23\times10^{-6}$  & $2.80\times10^{-7}$ & $7.01\times10^{-7}$ \\    
\hline
$10^5$& $1.12\times10^{-6}$ &$3.64\times10^{-7}$ &$2.17\times10^{-6}$& $4.48\times10^{-6}$ &$5.45\times10^{-8}$&$9.20\times10^{-7}$\\
\hline
$10^6$ & $3.51\times10^{-8}$ &$1.14\times10^{-8}$& $2.10\times10^{-6}$& $4.46\times10^{-6}$&$5.222\times10^{-8}$& $9.22\times10^{-7}$\\
\hline
$10^7$ & $1.11\times10^{-9}$ &$3.68\times10^{-10}$&$2.10\times10^{-6}$& $4.46\times10^{-6}$& $5.21\times10^{-8}$& $9.22\times10^{-7}$\\
\hline
$10^8$ & $2.18\times10^{-11}$ &$1.30\times10^{-11}$&$2.10\times10^{-6}$&$4.46\times10^{-6}$&  $5.22\times10^{-8}$& $9.22\times10^{-7}$\\
\hline
$10^9$ & $3.72\times10^{-10}$ &$3.37\times10^{-10}$&$2.10\times10^{-6}$& $4.46\times10^{-6}$&$5.22\times10^{-8}$&$9.25\times10^{-7}$\\
\hline
\end{tabular}
\end{center}
\end{flushleft}}
\caption{}
\label{tab:erroresd}
\end{table}

\newpage

\newpage

\newpage
\newpage

\newpage
\newpage

\newpage

\newpage

\newpage

\newpage

\newpage

\newpage

\clearpage
\newpage

\newpage

\clearpage

\clearpage
\newpage

\clearpage
\newpage

\clearpage

\clearpage

\clearpage

\clearpage

\clearpage

\clearpage

\clearpage

\clearpage

\clearpage

\begin{thebibliography}{}
\bibitem{Aref} Aref, H., 1984. Stirring by chaotic advection. J. Fluid Mech. 143, 1 ­ 21.

\bibitem{tpfl}  Khakhar DV, Ottino J.M. Fluid mixing (stretching) by time-periodic sequences for weak flows.
Physics of Fluids  29   11    3503-3505   (1986).
 
\bibitem{3dsfl}Dombre T., Frisch U., Greene J.M., Henon M., Mehr A., Soward A.M. Chaotic Streamlines in the ABC flows
Journal of Fluid Mechanics  167,    353-391   (1986).  

\bibitem{wiggins} Wiggins, S.: Chaotic Transport in Dynamical Systems, Springer-
  Verlag, New York, 1992

\bibitem{wigg} Wiggins, S.: Introduction to Applied Nonlinear Dynamical Systems and Chaos , Springer-Verlag, New York, 2003.

\bibitem{guck} Guckenheimer, J., Holmes, P.: Nonlinear Oscillations, Dynamical Systems, and Bifurcations of Vector Fields, Springer-
  Verlag, New York, 2002

\bibitem{malhotra} Malhotra, N. and Wiggins, S.: Geometric Structures, Lobe Dynamics, and Lagrangian Transport in Flows with Aperiodic Time Dependence, with Applications to Rossby Wave Flow, J. Nonlinear
  Sci., 8, 401­456, 1998.

\bibitem{haller1} Haller, G. and Poje, A.: Finite time transport in aperiodic flows,
  Physica D, 119, 3/4, 352­380, 1998.

\bibitem{langa1} Langa JA, Robinson JC, Suarez A. Stability, instability, and bifurcation phenomena in non-autonomous differential equations. Nonlinearity   15   (3)  887-903   (2002).
\bibitem{langa2}Langa JA, Robinson JC, Suarez A. Bifurcations in non-autonomous scalar equations. Journal
of Differential Equations.  221 (1) 1-35   (2006).
\bibitem{physrep} Mancho A. M., Small D., Wiggins S., A tutorial on dynamical systems concepts applied to Lagrangian transport in oceanic flows defined as finite time data sets: Theoretical and computational issues. Physics Reports 437, 3-4 (2006).

\bibitem{nlpg} Mancho, A. M., Small, D., Wiggins, S., 2004. Computation of hyperbolic and their stable and unstable manifolds for oceanographic flows represented
  as data sets. Nonlinear Process. Geophys. 11, 17 ­ 33.

\bibitem{jpo} Mancho A. M., Hern\'andez-Garc\'{\i}a E., Small D., Wiggins S., Fern\'andez V. Lagrangian transport through an ocean front in the North-Western Mediterranean Sea. Journal of Physical Oceanography 38, 6, 1222-1237 (2008).

\bibitem{aurell} Aurell E., Boffetta G., Crisanti A., Paladin G., Vulpiani A. Predictability in the large: an extension of the concept of Lyapunov exponent. J. Phys. A: Math. General, 30: 1-26, 1997.

\bibitem{haller} Haller, G. Distinguished Material surfaces and coherent structures in three-dimensional fluid flows. 
Physica D, 149: 248-277, 2001.

\bibitem{nese} Nese JM, Quantifying local predictability in phase space. Physica D 35: 237-250, 1989.

\bibitem{kayo} Ide, K., Small, D., Wiggins, S., 2002. 
Distinguished hyperbolic trajectories in time dependent fluid flows: analytical and computational approach for
   velocity fields defined as data sets. Nonlinear Process. Geophys. 9, 237 ­ 263.

\bibitem{yu} Ju, N., Small, D., Wiggins, S., 2003. Existence and computation of hyperbolic trajectories of aperiodically 
time-dependent vector fields and their   approximations. Int. J. Bif. Chaos 13, 1449 ­ 1457.

\bibitem{haller2} G. Haller. Exact theory of unsteady separation for two dimensional flows. Journal of Fluid Mechanics 512, 257-311 (2004)

\bibitem{eis} Eisenbach S., Friedrich R.,  Large-eddy simulation of flow separation 
on an airfoil at a high angle of attack and Re=10(5) using Cartesian grids. Theoretical and Computational Fluid Dynamics
22   (3-4) 213-225   (2008). 



\bibitem{coulliete} Coulliette, C., Wiggins, S., 2001. Intergyre transport in a wind-driven, quasigeostrophic double gyre: an application of lobe dynamics. Nonlinear   Process. Geophys. 8, 69 ­ 94.

 \bibitem{szeri} Szeri, A., Leal, L. G., and Wiggins, S.: On the Dynamics of 
Suspended Microstructure in Unsteady, Spatially Inhomogeneous   Two-Dimensional Fluid Flows, 
J. Fluid Mech., 228, 207­241,  1991.


\bibitem{physd} Mancho, A. M., Small, D., Wiggins, S., Ide, K.,2003. 
Computation of Stable and Unstable Manifolds of Hyperbolic Trajectories in Two-Dimensional, Aperiodically Time-Dependent Vectors Fields. Physica D 182, 188-222.

\bibitem{cf} Mancho, A. M., Small, D., Wiggins, S., Ide, K.,2006. 
A comparison of methods for interpolating chaotic flows from discrete velocity data.
Computers and Fluids 35, 416-428.

\bibitem{nr} W.H. Press, S.A. Teukolsky, W.T. Vetterling, B.P. Flannery, Numerical Recipes in Fortran 77, The Art of Scientific Computing, 2nd ed., Cambridge University Press, Cambridge, 1999.


\bibitem{dr}  Dritschel, D.G., Contour dynamics and contour surgery: numerical algorithms for extended, high-resolution modelling of vortex dynamics  in two-dimensional, inviscid, incompressible flows, Comput. Phys. Rep. 10 (1989) 77­146.

\bibitem{br} Branicki, M, Mancho, A. M., Wiggins, S. A Lagrangian description of transport associated with
 a Front-Eddy interaction: application to data from the North-Western Mediterranean Sea.
Preprint submitted for publication.

\bibitem{samelson} D. B. Chelton, M. G. Schlax, R. M. Samelson, and R. A. de Szoeke, Global observations of large oceanic eddies, Geophysical Research Letters 34 (2007), L15606.

\bibitem{haller3}  Haller G, An objective definition of a vortex, Journal of Fluid Mechanics
 525, 1-26   (2005).


\end{thebibliography}
\end{document}